\documentclass{emulateapj}
\usepackage{apjfonts}
\usepackage[]{natbib}
\usepackage{graphics}
     
\newcommand{\etal}{et al.}  
\newcommand{\per}{\ensuremath{^{-1}}}
\newcommand{\persq}{\ensuremath{^{-2}}}

\newcommand{\hal}{H\ensuremath{\alpha}}
\newcommand{\hbeta}{H\ensuremath{\beta}} 
\newcommand{\hst}{\emph{HST}}
\newcommand{\mstar}{\ensuremath{M_{\star}}}
\newcommand{\msun}{\ensuremath{M_{\odot}}}
\newcommand{\lsun}{\ensuremath{L_{\odot}}}
\newcommand{\kms}{km s\ensuremath{^{-1}}}

\newcommand{\mbh}{\ensuremath{M_\mathrm{BH}}}

\newcommand{\sigmastar}{\ensuremath{\sigma_\star}}

\newcommand{\msigma}{\ensuremath{\mbh-\sigmastar}}

\newcommand{\mgb}{Mg\ensuremath{b}}
\newcommand{\rosat}{\emph{ROSAT}}
\newcommand{\lbol}{\ensuremath{L_{\mathrm{bol}}}}

\newcommand{\ledd}{\ensuremath{L_{\mathrm{Edd}}}}
\newcommand{\dn}{\ensuremath{D_n(4000)}}
\newcommand{\hdeltaa}{H\ensuremath{\delta_A}}
\newcommand{\lstar}{\ensuremath{L^\star}}
\newcommand{\fwhmo}{\ensuremath{\mathrm{FWHM}_\mathrm{[OIII]}}}

\newcommand{\sigmao}{\ensuremath{\sigma_\mathrm{[OIII]}}}
\newcommand{\lothree}{\ensuremath{L_\mathrm{[OIII]}}}

\shorttitle{LOW-MASS SEYFERT 2 GALAXIES} 
\shortauthors{BARTH ET AL.}

\begin{document} 

\title{Low-Mass Seyfert 2 Galaxies in the Sloan Digital Sky Survey}

\author{Aaron J. Barth}
\affil{Department of Physics and Astronomy, 4129 Frederick
  Reines Hall, University of California, Irvine, CA 92697-4575}
\author{Jenny E. Greene\altaffilmark{1}}
\affil{Department of Astrophysical Sciences, Princeton
  University, Princeton, NJ, 08544}
\altaffiltext{1}{Hubble Fellow}
\and
\author{Luis C. Ho}
\affil{The Observatories of the Carnegie Institution of
  Washington, 813 Santa Barbara Street, Pasadena, CA 91101}

\begin{abstract}

We describe a sample of low-mass Seyfert 2 galaxies selected from the
Sloan Digital Sky Survey, having a median absolute magnitude of $M_g =
-19.0$ mag.  These galaxies are Type 2 counterparts to the Seyfert 1
galaxies with intermediate-mass black holes identified by Greene \& Ho
(2004).  Spectra obtained with the \emph{Echellette Spectrograph and
Imager} at the Keck Observatory are used to determine the central
stellar velocity dispersions and to examine the emission-line
properties.  Overall, the stellar velocity dispersions are low
($\sim40-90$ \kms), and we find 12 objects having $\sigmastar < 60$
\kms, a range where very few Seyfert 2 galaxies were previously known.
The sample follows the correlation between stellar velocity dispersion
and FWHM([\ion{O}{3}]) seen in more massive Seyfert galaxies,
indicating that the narrow-line FWHM values are largely determined by
virial motion of gas in the central regions of the host galaxies, but
the [\ion{O}{3}] emission lines exhibit a higher incidence of redward
asymmetries and double-peaked profiles than what is found in typical
Seyfert samples.  Using estimates of the black hole masses and AGN
bolometric luminosities, we find that these galaxies are typically
radiating at a high fraction of their Eddington rate, with a median
$\lbol/\ledd = 0.4$.  We identify one galaxy, SDSS
J110912.40+612346.7, as a Type 2 analog of the nearby dwarf Seyfert 1
galaxy NGC 4395, with a nearly identical narrow-line spectrum and a
dwarf spiral host of only $M_g = -16.8$ mag.  The close similarities
between these two objects suggest that the obscuring torus of AGN
unification models may persist even at the lowest luminosities seen
among Seyfert galaxies, below $\lbol = 10^{41}$ ergs s\per.
Spectropolarimetry observations of four objects do not reveal any
evidence for polarized broad-line emission, but SDSS
J110912.40+612346.7 has a continuum polarization significantly in
excess of the expected Galactic foreground polarization, possibly
indicative of scattered light from a hidden nucleus.  Forthcoming
observations of this sample, including X-ray and mid-infrared
spectroscopy, can provide new tests of the obscuring torus model for
active galaxies at low luminosities.

\end{abstract}


\section{Introduction}

The majority of active galactic nuclei (AGNs) are found in giant
galaxies with substantial bulges, and there is a dramatic drop in the
AGN fraction for late Hubble types \citep{hfs97} and for host galaxies
with stellar masses below $\sim10^{10}$ \msun\ \citep{kau03agn}.  This
trend is consistent with the generally accepted scenario in which
black hole growth and bulge growth are closely coupled, as expected
from the correlations between black hole mass and bulge mass
\citep{kr95}, and between black hole mass and stellar velocity
dispersion \citep[the \msigma\ relation;][]{fm00, geb00}.  The low AGN
fraction in late-type and low-mass galaxies arises from a combination
of factors.  Perhaps most important, the black hole occupation
fraction in low-mass galaxies is apparently below unity, as
demonstrated by the stellar-dynamical non-detections of central black
holes in the Local Group galaxies M33 \citep{geb01, mfj01} and NGC 205
\citep{val05}.  For those late-type galaxies that do contain a central
black hole, shallow central gravitational potential wells may lead to
a low efficiency for fueling black hole accretion.  A low-mass black
hole ($\lesssim10^6$ \msun), even if radiating at its Eddington rate,
cannot produce a very luminous AGN, and such objects can only be
readily detected as AGNs if they are fairly nearby.  In optical
surveys, dust extinction and blending with circumnuclear star-forming
regions can further hinder the detection of low-luminosity active
nuclei in late-type spirals.

If the \msigma\ relation continues toward low masses, then a black
hole of $10^6$ \msun\ would correspond to a host galaxy velocity
dispersion of $\sigmastar\approx60$ \kms\ \citep{tre02}.  Thus, to
explore the intermediate-mass regime for black holes, it is
particularly interesting to search for AGN host galaxies having
$\sigmastar<60$ \kms.  Such objects are known to exist, but they are
rare.  Prior to the Sloan Digital Sky Survey (SDSS), there were only
two nearby examples of dwarf galaxies known to contain Seyfert 1
nuclei: POX 52, a dE galaxy with $\sigmastar=36\pm5$ \kms\
\citep{ksb87, bar04}, and NGC 4395, an Sd-type spiral with a central
velocity dispersion of $\sigmastar\lesssim30$ \kms\ \citep{fs89,
fh03}.  The availability of SDSS has made it possible for the first
time to search systematically for more examples of AGNs with low-mass
black holes and small central velocity dispersions.  Greene \& Ho
(2004; hereinafter GH04) carried out the first such survey using Data
Release 1 of the SDSS, identifying 19 Seyfert 1 galaxies as likely
candidates for having $\mbh < 10^6$ \msun.  Although host galaxy
properties were not considered in the sample selection, the hosts for
this sample turned out to be galaxies of relatively low luminosity, on
average about 1 mag fainter than \lstar.  \citet{bgh05} measured
stellar velocity dispersions for many of the galaxies in this sample
and found that they are also low ($\sim35-80$ \kms) and fall close to
the local \msigma\ relation extrapolated toward lower masses.

Most previous searches for low-mass AGNs have concentrated on Type 1
(broad-lined) objects (Greene \& Ho 2004, 2007; Dong \etal\ 2006)
because the broad-line widths and AGN continuum luminosity can be used
to estimate the black hole mass.  However, at low luminosities, most
Seyferts are Type 2 objects \citep{hfs97, hao05b}, and a full
determination of the demographics of AGNs with low-mass black holes
must take into account the Type 2 population.  Furthermore, without
the glare of a bright AGN point source, the host galaxies of Seyfert
2s are more easily studied.  This can make it possible to examine the
central stellar populations and star formation histories of Seyfert 2
host galaxies in a level of detail that would be impossible for Type 1
AGNs \citep{kau03agn, hec04}.  Since searches for broad-lined AGNs
require both the flux and the width of a broad emission line (such as
\hal) to be above some detection threshold, surveys for Type 2 AGNs
based on measurements of narrow emission lines can potentially probe
lower AGN luminosities, and therefore may be sensitive to AGNs in
smaller host galaxies, or containing black holes of lower mass than
those that can be found in Seyfert 1 surveys.  Optical spectroscopic
surveys such as the Palomar survey \citep{hfs97} have previously
identified some examples of very low-luminosity Seyfert 2 nuclei in
late-type galaxies, such as the Sc galaxy NGC 1058, which has a
central velocity dispersion of only $31\pm6$ \kms\ \citep{bhs02}.
Other strategies for finding obscured AGNs, such as mid-infrared
searches for high-ionization coronal line emission from obscured
active nuclei \citep{sat07}, have contributed further evidence for
black holes in late-type spirals.  Since stellar-dynamical searches
for low-mass black holes are limited to very nearby galaxies (within a
few Mpc at best), AGN surveys remain the best way to examine the
demographics of these objects, and to determine the properties of the
host galaxies in which low-mass black holes form and grow.

This paper describes new observations of a set of Seyfert 2 galaxies
selected from Data Release 2 of the Sloan Digital Sky Survey
\citep[SDSS DR2]{aba04} to have host galaxies with relatively low
luminosities and small central velocity dispersions.  We select
galaxies spectroscopically identified as having Seyfert 2 nuclei
(rather than LINER or AGN/starburst transition or composite types), in
order to focus on objects having nuclear emission-line spectra that
are clearly and unambiguously dominated by an accretion-powered AGN.
These galaxies are in many respects Type 2 counterparts of the
low-mass Seyfert 1 sample found by GH04, although the two samples are
selected by very different criteria.  Most of the objects in both of
these samples are not small or faint enough to be considered dwarf
galaxies, but the objects described here are sub-\lstar\ galaxies that
occupy ranges in mass, luminosity, and central velocity dispersion
where very few Type 2 AGNs were known before SDSS.

We present measurements of the properties of these galaxies from
high-resolution spectra taken at the Keck Observatory.  The
small-aperture, high-resolution Keck spectra make it possible to
confirm the AGN classification of these galaxies, to detect weak
emission-line features and search for faint broad-line components, to
measure stellar velocity dispersions, and to fully resolve the
profiles of the narrow emission lines and measure accurate linewidths.
We discuss the properties of these galaxies and the Seyfert 1s from
the GH04 sample in the context of unified models of AGNs.  We
also describe a search for polarized broad-line and continuum emission
in four of these low-mass Seyfert 2 galaxies.  In this paper,
distance-dependent quantities are calculated assuming $H_0 = 72$ km
s\per\ Mpc\per, $\Omega_m$ = 0.3, and $\Omega_\Lambda = 0.7$.

\section{Sample Selection}

Our goal for this project was to identify Seyfert 2 galaxies having
low velocity dispersions, as Type 2 counterparts to the Seyfert 1
galaxies identified by GH04.  We initially searched the SDSS DR2
archives for narrow emission-line galaxies at $z<0.1$ with
emission-line ratios matching a Seyfert 2 classification (as described
below), and host galaxies with absolute magnitude fainter than $M_g =
-20$ mag.  In the GH04 Seyfert 1 sample, the host galaxy absolute
magnitudes ranged from $M_g = -17.8$ to $-20.6$ mag with a median of
$-18.9$ mag.  

The line-ratio selection was based on two criteria:
\begin{displaymath}
\log([\mathrm{O~III}]/\mathrm{H}\beta) > 0.61 /
    \{\log([\mathrm{N~II}]/\mathrm{H}\alpha) - 0.47\} + 1.19,
\end{displaymath}
and
\begin{displaymath}
[\mathrm{O~III}] / \mathrm{H}\beta >3,
\end{displaymath}
where [\ion{O}{3}] and [\ion{N}{2}] refer to the fluxes of the
[\ion{O}{3}] $\lambda5007$ and [\ion{N}{2}] $\lambda6583$ emission
lines, respectively.  The first criterion represents the ``maximum
starburst line'' of \citet{kew01}, which separates the \ion{H}{2}
region and AGN branches of a \citet{bpt81} ``BPT'' diagnostic diagram.
The location of this maximum starburst line was slightly modified in
later work by \citet{kau03agn} and \citet{kew06}, but our sample is
not affected by this change.  The second criterion was used to
identify high-excitation Seyfert galaxies and exclude LINERs or
objects with LINER/\ion{H}{2} transition-type spectra, following
traditional classification criteria \citep{hfs97}.

After we had begun some initial Keck observations, the SDSS DR2 galaxy
catalogs of G. Kauffmann and collaborators (the MPA/JHU catalogs),
based on work described by \citet{kau03sfr, kau03agn}, became
available to the public.\footnote{Updated versions of the MPA/JHU
catalogs described by Kauffmann \etal\ are available at
http://www.mpa-garching.mpg.de/SDSS~.}  The MPA/JHU AGN catalogs
include narrow emission-line galaxies falling above and to the right
of the maximum starburst line in the [\ion{O}{3}]/\hbeta\ vs.
[\ion{N}{2}]/\hal\ diagram, so their AGN samples contain both Seyferts
and LINERs.  For a description of their AGN sample selection and
classification criteria, see \citet{kau03agn}, which was based on the
SDSS DR1 sample.  We then used these catalogs to select further
candidates, and most of the objects we had initially selected from the
SDSS archives were included in the MPA/JHU DR2 AGN sample.

The MPA/JHU catalogs are an excellent resource for statistical studies
of nearby narrow-line AGN populations, because of the very large
sample size (33,589 AGNs in DR2 and 88,178 in the later DR4 catalog)
and the inclusion of additional data on the host galaxies including
stellar masses, stellar velocity dispersions, and the absorption-line
indices \hdeltaa\ and \dn, which are sensitive to the galaxy's star
formation history \citep[see][for details]{kau03sfr,kau03agn}.  To
ensure that our sample contained galaxies with well-determined AGN
classifications, we further selected only galaxies having measurements
of [\ion{O}{3}] flux with S/N$>10$ based on the DR2 emission line
catalog.  Several of the objects in our sample have also been
previously identified as Seyferts based on SDSS spectra by
\citet{hao05a}, but a literature search revealed that only two
galaxies in our sample (1440+0247 and 1032+6502) were identified as
AGNs prior to the SDSS.  A few cases of galaxies with possible broad
\hal\ emission lines, that otherwise met our selection criteria, were
included in the sample.  As described below, we used the Keck spectra
to test the reality of the broad-line emission.

\begin{figure}
\begin{center}
\plotone{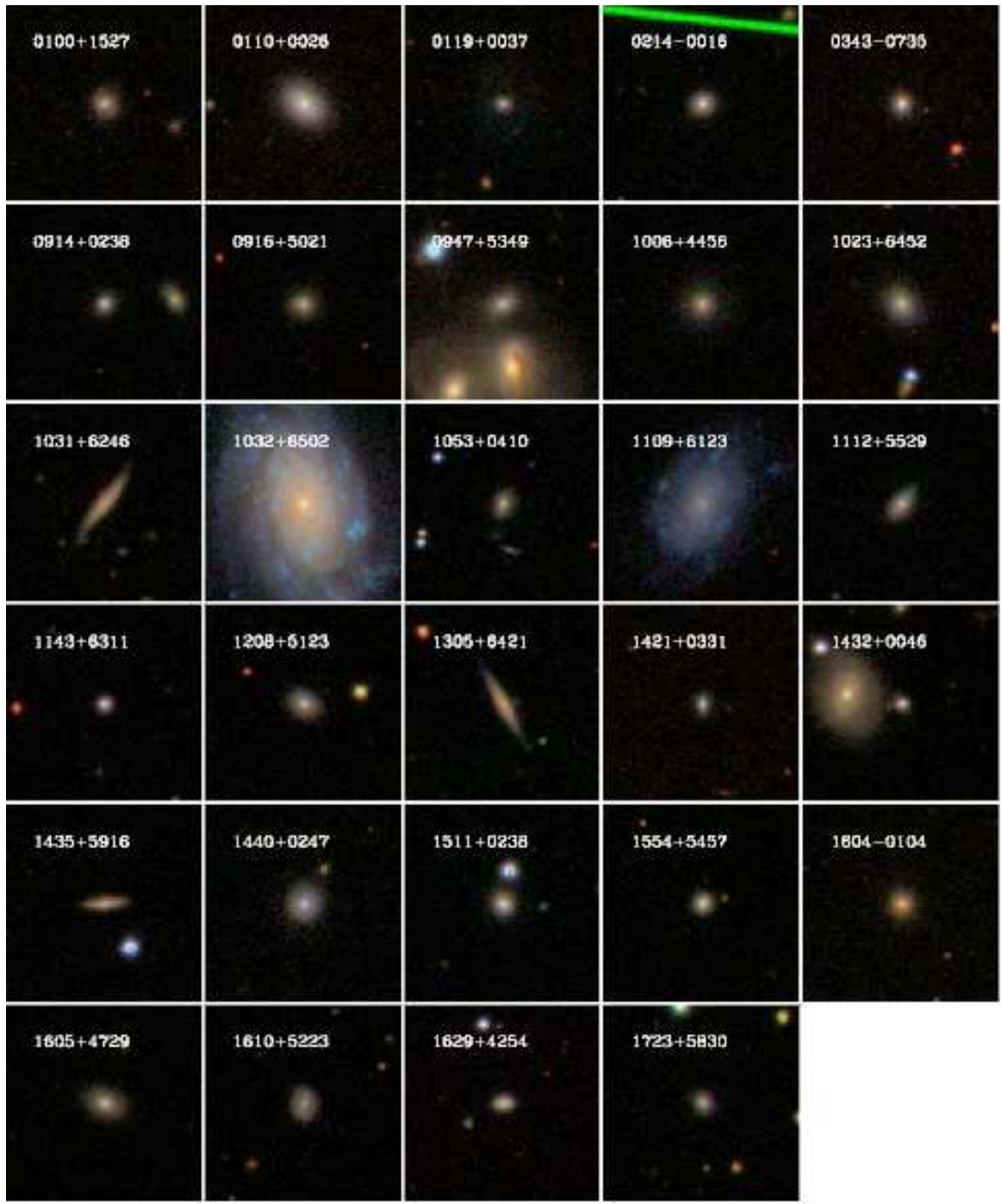}
\end{center}
\caption{SDSS color composite images of the Seyfert galaxy sample.
The image size is $1\arcmin\times1\arcmin$.  North is up and east is
to the left.
\label{images}}
\end{figure}

A total of 29 galaxies meeting these selection criteria were observed
at Keck.  The sample properties are described in Table \ref{sample},
and Figure \ref{images} displays SDSS images of each object.  The
redshifts range from $z=0.0056$ to 0.0712, with a median of 0.042.
Since the size of this Keck sample was determined by the available
telescope time and the practical need to observe objects distributed
over a range of right ascension, the set of objects in this study is
not a statistically complete sample, but they should be generally
representative of the small population of Seyfert 2 nuclei in
sub-\lstar\ host galaxies in SDSS.  In Table \ref{sample}, the
absolute $g$-band magnitudes are $K$-corrected to $z=0$ using
M. Blanton's \textit{kcorrect} IDL code \citep[version
4\_1\_4;][]{br07} and corrected for Galactic extinction using the dust
maps of \citet{sfd98}.

We note that the absolute magnitudes listed in Table \ref{sample} are
based on the SDSS Petrosian magnitudes \citep[see][]{str02}.  For most
objects in this sample, the Petrosian aperture magnitudes agreed to
within $\sim0.1$ mag with the ``model'' magnitudes based on the
best-fitting de Vaucouleurs or exponential model.  The magnitudes are
not corrected for internal extinction, since for most objects in the
sample there is insufficient morphological information to make such a
correction.  

For the objects selected from the MPA/JHU DR2 catalog, the host galaxy
stellar masses ranged from $\log(\mstar/\msun) = 8.1$ to 10.5, with a
median value of 9.8.  Although the MPA/JHU DR2 and DR4 AGN catalogs
contain a substantial number of AGN candidates with extremely low
stellar masses (i.e., below $10^9$ \msun), we found that the vast
majority of these objects do not satisfy the selection criteria given
above.  The MPA/JHU DR2 AGN catalog contained 90 objects with
$\log(\mstar/\msun) < 9.5$ and 17 with $\log(\mstar/\msun)<9.0$.
Inspection of the SDSS spectra of these objects revealed that most of
them have LINER or transition-type spectra, or very weak emission
lines with such low equivalent width that the AGN classification is
somewhat ambiguous.  There were also a few cases of nearby giant
galaxies (including NGC 5775 and NGC 3166) being listed in the AGN
catalog with very low host galaxy masses of $\log(M_\star/\msun) < 9$.
These appear to be objects for which the SDSS catalog magnitudes
include only the galaxy's nucleus, resulting in a severe underestimate
of the total stellar mass.  The situation for the DR4 AGN catalog is
similar: at $\log(M_\star/\msun) < 9$, there are 52 objects in this
catalog, most of which are either LINER or transition-type objects,
objects with emission lines of very low equivalent width, or large
nearby galaxies with incorrect total magnitudes.  Only a small
fraction of the objects in this mass range ($\lesssim10\%$) would
satisfy our Seyfert selection criteria.  Thus, the population
statistics of AGNs with extremely low-mass host galaxies in these
catalogs should be treated with some caution.

\section{Observations}
\label{sectionobservations}

\subsection{ESI Spectroscopy}

The spectroscopic observations described here were obtained with the
ESI spectrograph \citep{she02} on the Keck-II telescope.  Most of the
this sample was observed during 2005 May 16--17, with a few objects
observed during runs on 2003 November 23--24 and 2004 October 9--10.
We used the ESI echelle mode with a 0\farcs75-wide slit, resulting in
an instrumental dispersion of $\sigma_i \approx 22$ \kms\ and a total
wavelength coverage of 3850--11000 \AA.  The pixel scale in the
dispersion direction is 11.5 \kms\ pix\per.  The slit was held at a
fixed position angle during each observation, corresponding
approximately to the parallactic angle for the midpoint of the
exposure.  Exposure times for the galaxies ranged from 900 to 3600 s
(Table \ref{esidata}).  Seeing was mostly in the range
0\farcs7--1\farcs0.  One or more flux standards were observed during
each night, and several velocity template stars of spectral type
ranging from G8III to K4III, and also a few A0V stars, were observed
during twilight.  For wavelength calibration, exposures of HgNe, Xe,
and CuAr comparison lamp spectra were taken.

The spectra were bias-subtracted, flattened, and extracted with a
fixed extraction width of 1\arcsec.  Standard (not optimally weighted)
extractions were used, because optimal extraction routines tend to
truncate the peaks of emission-line profiles when the emission-line
regions have a different spatial extent from the stellar continuum.
The extracted spectra for the individual echelle orders were
wavelength-calibrated and flux-calibrated, and corrected for telluric
absorption by dividing by a normalized spectrum of a white dwarf star
observed on the same night.  Error spectra were also extracted and
propagated through the full sequence of calibrations.  Finally, the 10
echelle orders were combined into a single spectrum for each galaxy,
using a weighted average to combine the overlap regions between
orders.  Systematic offsets between the flux scales of adjacent orders
within the overlap regions were typically smaller than 2.5\%.  The
combined spectra were binned to a logarithmic wavelength scale with a
binsize of $\Delta[\log(\lambda/$\AA$)] = 1.665\times10^{-5}$.  Figure
\ref{spectra} displays the central portion of the ESI spectra.

\subsection{Spectropolarimetry}

We obtained spectropolarimetry observations of four objects using the
LRIS polarimeter \citep[LRISp;][]{gcp95} on the Keck I telescope on
the night of 2005 December 30, in clear conditions with 0\farcs7
seeing.  The instrumental setup consisted of a 400 lines mm\per\ grism
on the blue side of the spectrograph, covering 3200--5700 \AA\ at 1.09
\AA\ pixel\per, and a 600 lines mm\per\ grating on the red side,
covering 5500--8000 \AA\ at 1.28 \AA\ pixel\per.  A 1\arcsec-wide slit
was used, and the slit was oriented along the parallactic angle for
the midpoint of the exposure sequence.  Observations of each object
consisted of a sequence of four exposures with the half-wave plate
oriented at 0, 45, 22.5, and 67.5 degrees.  For three galaxies
(0119+0037, 0214-0016, and 0947+5349), one exposure set was taken, for
a total on-source exposure of one hour, and for the galaxy 1109+6123
we obtained two exposure sequences for a total of two hours of
integration.  Calibrations were performed using observations of
polarized and unpolarized standards \citep{mf70,ct90,tur90}, and flux
standard stars observed during twilight.  Data reduction and
polarization analysis followed the methods outlined by \citet{mrg88},
using a 3\arcsec-wide spectral extraction.  The null standard star HD
57702 was found to be unpolarized to within $0.08\%$ in both the blue
and red side extractions, in close agreement with other recent
measurements done with LRISp \citep{leo02}.

\section{Results and Discussion}

\subsection{Emission-Line Diagnostics}

Before measurement of the emission lines from the ESI spectra, we
applied a simple starlight subtraction procedure to remove the stellar
continuum from each spectrum. The fitting routine used a linear
combination of two stellar templates, a K giant and an A0V star, from
the library of stellar spectra observed on the same nights as the
galaxies.  A featureless, power-law continuum component was added, and
the continuum was velocity-broadened by convolution with a Gaussian
kernel and fitted to the galaxy spectrum using a Levenberg-Marquardt
minimization routine.  The free parameters in the fit included the
normalization of the two stellar spectra and the featureless
continuum, the stellar velocity dispersion, the power-law index of the
featureless continuum, and the reddening (assuming a Galactic
extinction law).  Spectral regions containing emission lines were
masked out in the fit.  This procedure was repeated for each galaxy
using several different late-type giant stars and A0V stars to find
the best-fitting combination.  While the starlight-subtracted spectra
of some galaxies still contained broad residual features or small
jumps at the boundaries between echelle orders, this procedure
generally yielded a good fit to the continuum shape and the strong
stellar features over the region $\sim4000-7000$ \AA, as illustrated
in Figure \ref{starsub}.  Although these fits give an estimate of the
stellar velocity dispersion, better measurements of the velocity
dispersions are determined from fits to smaller spectral windows in
individual echelle orders containing strong absorption-line features
(as described below), and we used these global fits primarily for
purposes of starlight subtraction to obtain the emission-line spectra.

The fluxes of the \hal, \hbeta, [\ion{O}{3}] $\lambda5007$,
[\ion{O}{1}] $\lambda6300$, [\ion{N}{2}] $\lambda\lambda6548, 6583$,
and [\ion{S}{2}] $\lambda\lambda6716, 6731$ emission lines were
measured by Gaussian fitting to the line profiles in the
starlight-subtracted spectra.  Each line was first fit with a single
Gaussian component, and if there were significant systematic
residuals, a second Gaussian component was added.  For the
[\ion{N}{2}] and [\ion{S}{2}] doublets, the wavelength separations
between each component were fixed to their laboratory values, and the
velocity widths of the corresponding components for each line were
held to be equal.  The flux ratio of the [\ion{N}{2}] $\lambda6583$
and 6548 lines was set to 2.96:1.

From inspection of the SDSS spectra of these galaxies, four objects
appeared to be likely or possible candidates for having broad \hal\
emission, and we used a single Gaussian component to fit the broad
line (Figure \ref{broadhal}).  Two of the broad \hal\ candidates were
the two objects that were not selected from the MPA/JHU catalog:
1032+6502 and 1440+0247.  In the case of 1032+6502, inspection of the
starlight-subtracted ESI spectrum clearly confirms the reality of the
broad \hal\ feature, and the broad component in the ESI data has FWHM
= $3020\pm33$ \kms.  The result of the model fit for 1440+0247 is
somewhat more ambiguous.  A fit using a double-Gaussian narrow-line
model left very weak residuals in the wings of \hal\, but the
residuals only extended in wavelength as far as the gaps between \hal\
and the [\ion{N}{2}] lines on either side.  Adding a broad \hal\
component marginally improved the quality of the fit, and the
best-fitting model had a broad component width of FWHM = $1040\pm20$
\kms.  Given the narrowness and low amplitude of this possible broad
component, it is not clear whether this represents emission from a
distinct broad-line region (BLR), and we consider it a tentative
detection at most.

The other two broad \hal\ candidates were selected from the MPA/JHU
catalog.  The first, 0110+0026 has an obvious and strong broad \hal\
emission line that is clearly visible in the SDSS spectrum and
confirmed in the ESI data, with FWHM = $5170$ \kms.  The other
candidate is 1629+4254; for this object both the SDSS and ESI spectra
show broad bases to all of the narrow emission lines.  Fitting the
lines with a narrow core and broad redshifted base for each narrow
line, we were able to reproduce the overall shape of the
\hal+[\ion{N}{2}] blend adequately.  Although the double-Gaussian
model does not perfectly reproduce the narrow-line profiles, the broad
bases are present on all of the forbidden lines and there does not
appear to be any need to include a separate broad \hal\ component in
the fit.  For the remaining objects, no broad-line component was
required to fit the \hal+[\ion{N}{2}] spectral region in the ESI
spectra.

Figure \ref{bptdiagrams} shows the location of the galaxies in this
sample (measured from the Keck ESI spectra) on line-ratio diagnostic
diagrams, as well as the GH04 Seyfert 1 sample and the nearby AGNs NGC
4395 and POX 52.  The AGN classifications are essentially unchanged
between the SDSS 3\arcsec-diameter fiber measurements and the Keck
$0\farcs75\times1\arcsec$ aperture extractions.  In the smaller Keck
aperture, only one object falls just slightly below our
[\ion{O}{3}]/\hbeta\ selection threshold.

In the diagram of [\ion{O}{3}]/\hbeta\ vs.\ [\ion{N}{2}]/\hal, the
galaxies in this sample, as well as the GH04 sample, deviate
systematically away from the main locus of SDSS Seyfert 2 galaxies,
toward lower values of [\ion{N}{2}]/\hal.  NGC 4395 and POX 52 also
lie in this same sparsely populated region of the diagram, around
log([\ion{N}{2}]/\hal)$\approx -0.5$ and
log([\ion{O}{3}]/\hbeta)$\approx1$.  Photoionization modeling by
\citet{kra99} and \citet{gro06} has shown that this region of the
diagnostic diagram is occupied by Seyferts having lower narrow-line
region (NLR) metallicity than in typical Seyfert nuclei; for
high-excitation AGNs, decreasing the metallicity moves the line ratios
systematically leftward on the diagram.  This region of the diagram is
nearly empty, relative to the main Seyfert branch, as a result of the
low incidence of active nuclei in low-metallicity galaxies.  In the
models of \citet{gro06}, the majority of Seyferts have NLR metallicity
between $\sim2$ and 4$Z_\odot$, while NGC 4395 and POX 52, and the
more extreme objects in our sample and the GH04 sample, fall in the
region of the diagram corresponding to $Z \approx Z_\odot$.  At still
lower metallicities, below $0.5Z_\odot$, the AGN and \ion{H}{2}
sequences overlap, and additional diagnostics would be needed to
distinguish AGN from star-forming nuclei.  However, we note that the
multicomponent photoionization models for NGC 4395 described by
\citet{kra99} found a good match to the narrow-line spectrum for a
metallicity of $0.5Z_\odot$ overall and an N/H ratio of 1/6 solar.
Regardless of the exact value of the metal abundances in these
galaxies, the work of \citet{gro06} clearly demonstrates that AGN
selection based on low host galaxy mass efficiently identifies AGNs
with lower than average metallicity, and our sample follows this
trend.  In our sample, the object that deviates most from the normal
Seyfert branch in the BPT diagram is 1109+6123, which is also the
faintest and least massive galaxy in the sample.

We searched for high-ionization coronal line emission in the spectra,
as additional indicators of AGN activity.  The galaxy 1440+0247 shows
a range of coronal lines including [\ion{Fe}{6}] $\lambda5177$,
[\ion{Fe}{7}] ($\lambda5158$, 5720, and 6087 \AA), [\ion{Fe}{10}]
$\lambda6374$, and [\ion{Fe}{11}] $\lambda7802$ (Figure
\ref{1440lines}).  Its spectrum is extremely similar to that of NGC
4395 \citep{fs89, kra99} aside from the relative weakness of the broad
\hal\ emission.  The Type 1 object 1032+6502 also has weak
[\ion{Fe}{7}] and [\ion{Fe}{10}] emission.  Among the remaining Type 2
objects in the sample, these high-excitation Fe features are seen in
only one object, 0214$-$0016.  The LRISp spectra extend farther to the
blue than the ESI data, and three of the four LRISp targets
($0214-0016$, 0947+5349, and 1109+6123) have [\ion{Ne}{5}]
$\lambda3426$ emission in their spectra.

Since the individual ESI spectra have rather low S/N at the blue and
red ends, we created a composite Seyfert 2 spectrum by summing the
individual spectra (excluding the three objects with definite or
possible broad \hal\ emission), corrected for Galactic extinction and
weighted by the S/N of each spectrum.  Figure \ref{composite} shows
the composite spectrum, with the starlight subtraction procedure
applied as described above.  The best-fitting K+A spectral model for
the composite spectrum shows deep Balmer absorption, and reveals
emission lines that are difficult to detect clearly in many of the
noisy individual spectra, including [\ion{Ne}{3}]
$\lambda\lambda3868,3967$, [\ion{Fe}{7}] $\lambda6086$, and the
[\ion{S}{3}] $\lambda\lambda$9069, 9532 lines at the red end.  We
measured emission line fluxes from the composite spectrum; when
necessary, in regions where the starlight subtraction left significant
residuals a spline was first fitted to the local continuum and
subtracted in order to flatten the continuum.  Emission-line fluxes
measured from the composite spectrum are given in Table
\ref{compositelines}.

\subsection{[\ion{O}{3}] Profiles}
\label{forbiddenlines}

Following methods similar to those described by \citet{gh05}, we
measured the linewidths of strong emission lines from fits of single
or double-Gaussian models to each line.  We focus on the [\ion{O}{3}]
$\lambda5007$ emission line since this is the strongest forbidden line
in the spectra and it is uncontaminated by underlying stellar
absorption or blending with other emission lines.  For three galaxies
in the sample, the [\ion{O}{3}] profiles were adequately modeled by
single Gaussian components with no strong systematic residuals.  In
all other cases, two Gaussian components were used to model each
narrow emission line.  This type of line-profile decomposition is
often interpreted in terms of a core component dominated by virial
motions and a wind-dominated wing component \citep[e.g.,][]{gh05},
although some objects in our sample have more complex profiles that do
not fit this simple qualitative description.  Figure \ref{o3} shows
the [\ion{O}{3}] $\lambda5007$ profiles for the sample.

The [\ion{O}{3}] profiles exhibit a greater degree of diversity than
seen in previous SDSS samples.  It has long been known that blueward
asymmetries on [\ion{O}{3}] profiles are much more common than redward
asymmetries \citep[e.g.,][]{om86}.  In the large sample of 1749 SDSS
Seyfert 2 spectra studied by \citet{gh05}, most of the narrow-line
profiles were well described by a narrow core component plus a
low-amplitude, blueshifted ``wing'' component.  The wing component was
found to be redshifted relative to the core in only 6\% of the
objects.  In contrast, our sample seems to have an unusually large
fraction of objects with redward asymmetries.  Eight objects, or 28\%
of the sample, have red-asymmetric profiles, in the sense of having a
single dominant peak with an excess of flux on the red side of the
peak relative to the blue side.  Some other objects (such as 1112+5529
and 1432+0046) have profiles that are well fit by the double-Gaussian
model but for which the model components cannot be individually
labeled as core or wing components, because the broader component
contains the majority of the line flux and the velocity difference
between the two components is small.  Either blueward or redward
asymmetries can be interpreted as resulting from radial motion (inflow
or outflow) of clouds combined with the presence of dust, either
within the clouds themselves or in a surrounding medium.  In the more
common situation of blueward asymmetries, a likely explanation is
outflowing motion of clouds in a dusty medium, in which the redshifted
clouds on the far side of the galaxy nucleus are behind a larger
absorbing column of dust \citep[e.g.,][]{heck81}.  However, there are
possible geometric arrangements of an outflow and dust screen that
would lead to a redward asymmetry as well, if the outflow direction is
not restricted to be perpendicular to the plane of the dust screen.
Alternatively, if the dust is contained within the NLR clouds
themselves, then the emission from a cloud would be strongest in the
direction facing the central ionizing source, and in an outflow we
would have a more direct view of the ionized faces of the redshifted
clouds on the far side of the nucleus, rather than the ionized faces
of the blueshifted clouds, leading to a redward asymmetry on the line
profiles.  As discussed by \citet{whit85}, the line profiles do not
contain sufficient information to uniquely determine the kinematic
state of the NLR or the distribution of dust within it.
 
More surprising is the detection of double-peaked forbidden-line
profiles in three objects.  Two galaxies, 0343$-$0735 and 1006+4456,
have fairly symmetric double-peaked [\ion{O}{3}] profiles with
well-resolved peaks.  The peak-to-peak velocity separation of the two
components in these objects is 115 and 275 \kms, respectively.  The
most spectacular example is 1629+4254, which contains a narrow core
component having FWHM = 78 \kms\ and a very broad, redshifted wing
component with a peak velocity separation of 223 \kms\ and FWHM = 589
\kms.  Additionally, a marginally resolved second peak is visible on
the red shoulder of the [\ion{O}{3}] line profile in 0914+0238.  For
comparison, the \citet{gh05} study found double-peaked profiles in
only about 1\% of SDSS Seyfert 2 spectra.  The relatively high
incidence of double-peaked profiles in this sample likely results from
a combination of the high spectral resolution of ESI, the smaller
spectroscopic aperture of the Keck observations, and the low intrinsic
linewidths for the individual velocity components in these low-mass
galaxies.  In the SDSS spectra, velocity splittings much smaller then
the instrumental FWHM of $\sim170$ \kms\ would be unresolved, and only
one of these four objects (1006+4456) exhibits double-peaked lines in
the lower-resolution SDSS data.  The SDSS spectrum for 1629+4254 shows
a single-peaked but strongly red-asymmetric [\ion{O}{3}] profile.

Double-peaked narrow emission lines can arise from a disklike NLR,
from outflows, or possibly from a binary AGN with two distinct NLRs
\citep[e.g.,][]{zhou04}.  While a fast outflow is the most likely
explanation for the extremely broad and asymmetric profile seen in
1629+4254, the galaxies $0343-0735$ and 1006+4456 have more symmetric
double-peaked [\ion{O}{3}] profiles suggestive of a disklike emitting
region.  The possibility of a disklike NLR would be difficult to test
directly by spatially resolved kinematics or morphology, however,
since the diameter of such a disk would be too small to resolve
directly in these galaxies.

The linewidths for [\ion{O}{3}] were characterized by two different
quantities: the line dispersion (that is, the second moment of the
line profile) and the full-width at half-maximum.  Given the limited
S/N in the profile wings in some objects, we chose to measure the
linewidth parameters from the fitted model profile rather than from
the data itself, following methods similar to those used by
\citet{gh05}.  The line dispersion is given by
\begin{displaymath}
\sigma_\mathrm{line}^2 = \left(\frac{c}{\lambda_0}\right)^2 
\frac{\int{(\lambda - \lambda_0)^2 f_\lambda
    d\lambda}}{\int{f_\lambda d\lambda}},
\end{displaymath}
where $\lambda_0$ is the flux-weighted mean wavelength of the emission
line.  The FWHM was measured as the full velocity width at half of the
peak flux density of the modeled line profile.  In the case of a
Gaussian profile, $\sigma=$FWHM/2.35.  For line profiles having the
prototypical core/wing structure, the FWHM is primarily sensitive to
the width of the narrow core component, and the ratio of $\sigma$ to
$(\mathrm{FWHM}/2.35)$ gives a measure of the relative prominence of
the line wings.

The instrumental dispersion was removed by subtracting it in
quadrature from the measured line dispersions, according to
$\sigma_\mathrm{line}^2 = \sigma_\mathrm{obs}^2 -
\sigma_\mathrm{inst}^2$, and for the FWHM values we subtracted
($2.35\sigma_\mathrm{inst}$) in quadrature from the measured values.
Uncertainties on FWHM and $\sigma$ were determined by Monte Carlo
simulations.  Starting with the single or double-Gaussian fit
parameters for each galaxy, we created 1000 realizations of the line
profile, drawing the amplitude, width, and centroid of each Gaussian
component randomly from a normal distribution based on the fitting
uncertainties from the original profile fit.  Noise was then added to
match the S/N of the original spectrum, and the simulated profile was
fitted in the same manner as the original spectrum.  The uncertainty
on the line dispersion or FWHM was then taken to be the standard
deviation of linewidth values measured from the set of simulated
profiles.  The results are listed in Table \ref{esidata}.  The FWHM
values for the [\ion{S}{2}] lines, measured in the same manner, are
also listed.

As expected for galaxies of relatively low mass, the narrow emission
lines in these galaxies are much narrower than those of more luminous,
classical Seyfert 2 galaxies.  Excluding the three strongly
double-peaked objects, the median values of \fwhmo\ and \sigmao\ are
132 and 87 \kms, respectively, and the narrowest-lined object,
1109+6123, has \fwhmo\ = 66 \kms.

\subsection{Stellar Velocity Dispersions}
\label{veldisp}

Stellar velocity dispersions (\sigmastar) were measured from the ESI
spectra by direct fitting of velocity-broadened stellar templates,
following the method described by \citet{bhs02}. Although the
starlight subtraction procedure described above in
\S\ref{forbiddenlines} yielded an estimate of the stellar velocity
dispersion, the starlight subtraction fits were performed over the
entire wavelength range of the data. For accurate measurement of
velocity dispersions it is critical to perform the measurements over a
restricted wavelength range containing strong stellar absorption
features.  It is also critical to mask out wavelengths containing
narrow emission lines so that they do not affect the dispersion
measurements.  To eliminate any potential issues of mismatch in
spectral resolution between echelle orders, we carried out the
velocity dispersion measurements using spectra from a single echelle
order.

Measurements were performed independently over two wavelength ranges,
as illustrated in Figure \ref{dispersions}: a blue region (5130--5470)
including the strong Mg$b$ and Fe 5270 lines, and the \ion{Ca}{2}
triplet region (8450--8700 \AA).  When visible in the spectra or in
the fit residuals, emission lines such as [\ion{N}{1}]
$\lambda\lambda5199,5201$ were masked out from the fits.  For the
velocity dispersion measurements, the model fits consisted of a
template star broadened by a Gaussian in velocity, diluted by addition
of a featureless continuum (modeled by a quadratic polynomial in this
case), and multiplied by another quadratic polynomial that allows for
reddening or systematic flux calibration differences between the
galaxy and template star observations.  The velocity dispersion was
measured using template stars observed on the same night as each
galaxy.  Uncertainties on the velocity dispersions were determined as
the sum in quadrature of the formal fitting errors from the
best-fitting model and the standard deviation of the velocity
dispersions found from all available templates.  The final velocity
dispersion values listed in Table \ref{esidata} are the weighted
averages of the red and blue measurements.  As expected for galaxies
of low stellar mass, the velocity dispersions for this sample are low,
ranging from 39 to 94 \kms\ with a median value of 63 \kms, and with
12 objects having $\sigmastar<60$ \kms.  For two objects, 1053+0410
and 1109+6123, the continuum S/N was too low to permit a measurement
of $\sigmastar$.

These measurements give the dispersion of the line-of-sight velocity
profile for the integrated galaxy light falling in the ESI slit and
extraction aperture, and represent a combination of light from bulges
or pseudobulges, disks, and possibly central star clusters.
Higher-resolution images would be needed in order to determine the
relative contributions of starlight from the various structural
components.  Using bulge-disk decompositions of the GH04 Seyfert 1
sample from \emph{Hubble Space Telescope} (\hst) imaging,
\citet{ghb08} found that the measured dispersions from \citet{bgh05}
were dominated by bulge light for those systems with detected bulges,
but that in some host galaxies without detected bulges the measured
dispersion was likely dominated by disk starlight.  In either case,
the fact that the GH04 Seyfert 1 galaxies fall close to the \msigma\
relation extrapolated to low masses means that the central stellar
velocity dispersion measured in a $\sim1\arcsec$ aperture, regardless
of its origin, is still reasonably well correlated with black hole
mass for these low-mass AGNs.

\subsection{Comparison between Stellar and Gas Linewidths}

With measurements of both stellar velocity dispersions and
forbidden-line widths, we can examine whether the correlation between
\sigmastar\ and [\ion{O}{3}] linewidth, well established for typical
Seyfert 2 galaxies \citep{nw96}, continues to hold for these
low-dispersion Seyferts.  This issue has attracted considerable recent
attention as a result of the potential utility of the [\ion{O}{3}]
linewidth as a substitute for \sigmastar\ in studies of the \msigma\
relation for certain subclasses of AGNs for which \sigmastar\ can be
difficult or impossible to measure directly
\citep{wl01,shi03,gm04,bz04,sal07,byz06}.  For a large sample of
nearby Seyferts (mostly of Type 2), \citet{nw96} found that the mean
value of $\fwhmo/(2.35\sigmastar)$ was very close to unity, but that
for individual objects there was considerable scatter in the
relationship between stellar and gas linewidths.  Using SDSS spectra
of a large sample of low-redshift Seyfert 2 galaxies, \citet{gh05}
confirmed this general trend. Additionally, they found that Seyfert 2
galaxies tend to have an ``excess'' emission-line velocity dispersion,
over and above the stellar velocity dispersion, that increases as a
function of $\lbol/\ledd$, and that the [\ion{O}{3}] ``core''
component width (based on a double-Gaussian decomposition) is a better
predictor of \sigmastar\ than the dispersion of the full [\ion{O}{3}]
profile.  These trends are consistent with a scenario in which the
linewidth of the NLR core component is primarily set by virial
motion of gas in the host galaxy's bulge, while the increasing excess
linewidth with Eddington ratio reflects an increasing prevalence of
winds or outflows in highly accreting objects.  Since the SDSS spectra
themselves cannot be used to measure stellar or emission-line widths
for features with $\sigma \lesssim 70$ \kms, our sample allows us to
extend these relationships to galaxies with lower velocity
dispersions.

Figure \ref{linewidths} shows the comparison between stellar and
[\ion{O}{3}] linewidths, excluding the two galaxies for which
\sigmastar\ could not be measured as well as the three objects with
strongly double-peaked [\ion{O}{3}] profiles.  We examine both
\sigmao, which is sensitive to the profile wings, as well as \fwhmo.
In our sample, the distinction between core and wing
components is not always obvious from the line-profile fits, and we
cannot uniformly select a core component from the double-Gaussian
profile fits, but the FWHM values essentially reflect the width of the
line core. The measured values of \sigmao\ are systematically higher
than \sigmastar, while the values of FWHM([\ion{O}{3}])/2.35 track
\sigmastar\ much more closely.  To quantify this comparison, following
\citet{nw96} and \citet{gh05}, we measure the quantities $\Delta W =
\log(\fwhmo)- \log(2.35\sigmastar)$, and $\Delta\sigma = \log\sigmao -
\log \sigmastar$.  For our sample, we find a mean value of
$\langle\Delta\sigma\rangle = 0.14\pm0.15$, or alternatively
$\langle\sigmao/\sigmastar\rangle = 1.48\pm0.56$.  This is very
similar to the excess [\ion{O}{3}] linewidth in higher-luminosity
samples: \citet{gh05} found $\langle\sigmao/\sigmastar\rangle =
1.43\pm0.75$ for Seyfert 2 galaxies overall in the MPA/JHU DR2 catalog
with $\sigmastar>70$ \kms.  On the other hand, the \fwhmo\ values do
not reveal any strong excess linewidth for the core component.  The
mean value of $\Delta W$ for our sample is $\langle \Delta W \rangle =
-0.05 \pm 0.12$, indicating that on average the [\ion{O}{3}] FWHM
values track the stellar velocity dispersion (albeit with substantial
scatter), similar to results found by \citet{nw96} for more massive
Seyfert galaxies.  The [\ion{S}{2}] lines show a similar trend
relative to the stellar velocity dispersions, with $\langle \Delta
W_\mathrm{[SII]} \rangle = -0.05 \pm 0.15$.

Overall, these results demonstrate that the relationships between
[\ion{O}{3}] linewidth and stellar velocity dispersion continue to
hold for Seyfert 2 galaxies with stellar velocity dispersions of
$\sim40-80$ \kms, and that the \fwhmo\ values are primarily driven by
virial motion in the NLR.  Nevertheless, the excess linewidth seen in
the [\ion{O}{3}] line wings, and in the \sigmao\ values, indicates
that even for these relatively low-luminosity, low-mass AGNs,
nongravitational motion (presumably outflows driven by the central
engine) does play a significant role in the NLR kinematics.

\subsection{Polarization}

The spectropolarimetry results for the wavelength region surrounding
the \hal\ emission line in the red-side spectra are shown in Figure
\ref{polplot}.  Each panel displays the total flux spectrum, the
fractional polarization $p$ given as the ``rotated Stokes parameter''
\citep{mrg88}, and the Stokes flux, which is given by $p \times
f_\lambda$.  The clearest signature of a hidden broad line region
would be a broad ``bump'' in the $p$ spectrum at the \hal\ emission
line, which results in a Stokes flux spectrum resembling that of a
Seyfert 1 galaxy \citep{am85}.  Continuum polarization is another
possible sign of a hidden active nucleus, but some care must be taken
to distinguish it from polarization by transmission through aligned
dust grains, either in the host galaxy or of Galactic origin.

In the continuum near \hal, the S/N in the total flux spectra ranges
from $\sim40$ to 160 per pixel for these observations (Table
\ref{poldata}).  Since the spectra are strongly dominated by
starlight, only relatively strong polarized broad lines ($p \gtrsim
1-2\%$ above the continuum level) could potentially be detected in
observations of this quality, and none of these four objects exhibits
any obvious polarization feature at \hal.  The significance of any
possible features in the Stokes $q$ or $u$ spectra was quantified
using a method similar to that described by \citet{bfm99}.  For each
object, we fit quadratic polynomial models to the continuum in the
red-side $q$ and $u$ spectra, excluding a 150-\AA\ wide region around
\hal\ from the fit.  Over this 150-\AA\ wide region centered on \hal,
the excess polarization in the $q$ Stokes parameter relative to the
continuum polarization is then $\delta_q = \langle q_\mathrm{data} -
q_\mathrm{model} \rangle$.  To evaluate the uncertainty on this
measurement, we calculated the same quantity over a large number of
randomly placed 150-\AA\ wide bins in the continuum regions of the
spectrum over $5600-7600$ \AA.  The standard deviation of these
residual differences was taken to be the $1\sigma$ uncertainty
$\epsilon_q$ for the measurement at \hal.  This generally exceeds the
error level from the propagated photon-counting statistics and
detector readnoise because of small but unavoidable systematic errors
in spectral extractions such as errors in interpolating fractional
pixel values at the edges of the extraction apertures for spatially
extended objects \citep{bfm99}.  These errors result in broad
systematic wiggles in the $p$ spectra, as seen in Figure
\ref{polplot}.  The significance of any polarization excess at \hal\
relative to the fitted continuum level is then $\Delta_q = \delta_q /
\epsilon_q$, and similarly for the $u$ parameter.  The measurements
confirm that no significant \hal\ line polarization is detected.  The
maximum deviation from the continuum polarization level is only at the
$2\sigma$ level for the $u$ Stokes parameter in SDSS 0947+5349.  For
all other objects the Stokes parameters at \hal\ have $|\Delta_q| < 1$
and $|\Delta_u| < 1$.

The continuum polarization and polarization angle were measured over
wavelength bins $4400-5500$ \AA\ on the blue side and $6000-7500$ \AA\
on the red side.  All measurements were performed on the Stokes
parameter spectra, with the results converted to $p$ and $\theta$ as
the final step.  The results are reported in Table \ref{poldata},
along with the Galactic reddening $E(B-V)$ for these objects
\citep{sfd98} and the maximum expected level of Galactic foreground
polarization, $p_\mathrm{max} (\%) = 9.0 \times E(B-V)$ \citep{smf75}.
For the null standard HD 57702, we find blue and red-side
polarizations of $(0.07\pm0.01)\%$ and $(0.08\pm0.01)\%$,
respectively, so any systematic instrumental polarization should be at
or below this level.  The galaxies' continuum polarizations are $0.2$
to $0.76\%$, significantly larger than that of the null standard.

For three of the objects, the continuum polarization is only
marginally higher (by $\lesssim 0.2\%$) than the maximum expected
Galactic foreground polarization.  The only object exhibiting a
continuum polarization signal significantly above the Galactic
foreground level is SDSS 1109+6123, which has $p=(0.76\pm0.10)\%$ and
($0.72\pm0.10$)\% on the red and blue sides, respectively, while the
Galactic foreground $p_\mathrm{max}$ along this sightline is only
$0.1\%$.  Thus, the detected polarization almost certainly arises
within the host galaxy, but the limited S/N and lack of any polarized
line features does not permit a clear interpretation of its physical
origin.  Electron scattering or dust scattering of a hidden AGN
continuum could be responsible for the observed polarization.  If the
electron scattering mirror were located interior to the narrow-line
region then the narrow emission lines should have a lower polarization
than the surrounding continuum.  This is not observed in our data, but
would be testable with higher S/N observations.

Polarization by transmission through foreground dust in the host
galaxy is an alternative possibility, and is the dominant polarization
mechanism in some low-luminosity AGNs with high internal extinction
such as NGC 3718 \citep{bfm99}.  For SDSS 1109+6123, the Balmer
decrement \hal/\hbeta\ is $2.94\pm0.09$ in the SDSS spectrum and
$2.87\pm0.10$ in the LRISp spectrum, consistent with the Case B
recombination value of 2.85 and smaller than the typical AGN
narrow-line region value of $\sim3.1$ \citep[e.g.,][]{of06}, so the
reddening toward the NLR is too low to be determined accurately.  If
the dust transmission polarization within the host galaxy follows the
Galactic relationship between reddening and maximum polarization,
$p_\mathrm{max} = 9.0\times E(B-V)$ \citep{smf75}, then the observed
polarization would require $E(B-V) \gtrsim0.08$ mag, much higher than
the Galactic reddening of $E(B-V) = 0.011$ mag \citep{sfd98}.  For
Case B recombination this reddening would modify the observed Balmer
decrement to \hal/\hbeta$\gtrsim3.07$.  Given the uncertainties on the
measured Balmer decrements, we cannot fully rule out the possibility
that the observed continuum polarization is dominated by dust
transmission polarization within the host galaxy.  Scattered light
from a hidden AGN remains an intriguing possibility for this object,
but very long spectropolarimetry exposures would be needed in order to
achieve sufficient S/N to place interesting limits on the presence of
polarized broad \hal\ emission.

Nondetections of polarized emission lines or continuum in Seyfert 2
galaxies are unfortunately difficult to interpret; possible
explanations include lack of a BLR, lack of a suitable scattering
mirror, foreground dust absorption covering the host galaxy's nucleus,
or contamination by starlight that dilutes the strength of a
scattered-light polarization signal.  In starlight-dominated AGNs with
hidden BLRs, the broad-line polarization is often $\lesssim1\%$ above
the continuum polarization \citep{bfm99,mor00}, and the S/N of these
observations is too low to detect such weak polarized line features.

\subsection{Radio and X-ray Counterparts}

We searched for counterparts to these sources in the online source
catalogs of the FIRST \citep{bwh95} and NVSS \citep{con98} radio
surveys, and the \rosat\ All-Sky Survey \citep{vog99}.  Only two
objects were matched with sources in the radio catalogs: 1629+4254 is
detected in the FIRST survey with $f_\nu = 1.79$ mJy, and 1032+6502 is
an NVSS source with a flux density of 7.6 mJy.

The only two objects in this sample detected in the \rosat\ All-Sky
Survey are the two previously known AGNs with weak broad \hal\
emission, 1032+6502 and 1440+0247.  In both cases, the association
with the \rosat\ source has previously been noted: 1032+6502
\citep{bol92,mhh96} with a soft X-ray flux of $6.2\times10^{-13}$ ergs
cm\persq\ s\per, and 1440+0247 with a flux of $2.8\times10^{-13}$ ergs
cm\persq\ s\per\ in the \rosat\ 0.1--2.4 keV band \citep{and03}.

The lack of additional radio and X-ray detections for other objects in
this sample is not surprising.  Among the Type 1 objects found by
GH04, only one was detected in deep VLA imaging at 6 cm, and the
entire GH04 sample is extremely radio quiet \citep{ghu06}.  Similarly,
the GH04 sample is faint in X-rays, with only 30\% of the objects
detected in the \rosat\ All-Sky Survey.  If the Type 2 objects in our
sample are intrinsically similar objects with hidden broad-line
regions that are obscured either by a compact torus or by larger-scale
dust lanes in the host galaxy, the soft X-ray flux would likely be
strongly attenuated as well.  Deep \emph{XMM-Newton} observations of a
few objects in this sample have recently been obtained in order to
search for nuclear X-ray sources and to constrain the amount of
obscuring material toward the nuclei; the results of these
observations will be presented in a forthcoming paper.

\subsection{Host Galaxy Morphologies and Stellar Populations}

The host galaxies of AGNs with low-mass black holes exhibit a
surprising range of morphologies.  NGC 4395 is an essentially
bulgeless, late-type Sd spiral \citep{fs89}, and serves as the best
demonstration that black holes can form in some galaxies that lack a
substantial central bulge.  On the other hand, the structure of POX 52
is best described as a dwarf elliptical \citep{bar04, tho08}, although
with a host galaxy absolute magnitude of $M_V=-17.6$ and S\'ersic
index of $\sim4$ it is both more luminous and more centrally
concentrated than typical dE galaxies \citep{bj98}.  All of the host
galaxies in the GH04 Seyfert 1 sample are more luminous than these two
nearby objects, but analysis of \hst\ images shows that they also
exhibit a range of morphologies, with about 60\% being disk dominated
galaxies that contain a bulge or pseudobulge, and the remainder being
compact galaxies with spheroidal morphologies similar to POX 52
\citep{ghb08}.

From examination of the SDSS images for this sample, about half of the
objects appear to have definite or likely disks, while the remaining
objects apear compact without obvious disks.  These may be objects
with POX 52- type host galaxies, but higher-resolution imaging would
be needed to determine the morphological types.  The SDSS
concentration index $C$, defined as the ratio of the radii containing
90\% and 50\% of the total galaxy light in the $r$ band, gives a rough
quantitative measure of morphology \citep{str01}, as there is a
general separation between early-type, bulge-dominated galaxies with
$C\gtrsim2.6$ and late-type, disk-dominated galaxies with $C<2.6$.
The concentration values for our sample, taken from the MPA/JHU
catalog and SDSS archives, are listed in Table \ref{sample}.  The
median concentration for our sample is 2.59, only slightly lower than
the median concentration for the full AGN sample in the DR2 MPA/JHU
catalog overall, and the distribution of concentration values is
consistent with a mixed population including some elliptical or
spheroidal hosts.  Figure \ref{cmd} shows a
color-magnitude diagram for the sample, using the extinction-corrected
and $k$-corrected SDSS photometric data in the $g$ and $r$ bands, and
compared with standard galaxy colors from \citet{fuku95}.  The hosts
span a wide range in color, corresponding to typical colors ranging
from those of ellipticals to late-type spirals, with all but one
object having $g-r$ between 0.4 and 0.8.  

Only one galaxy in this sample, 1109+6123, appears to be a promising
candidate for a very late-type, nearly bulgeless ($\sim$Sd) disk
galaxy similar to NGC 4395 (see $\S$\ref{individualobjects}).  This
late-type disk galaxy is a notable outlier in this sample, both in
terms of its low luminosity ($M_g = -16.8$ mag) and its blue color
($g-r=0.28$ mag).  Even with SDSS, there are remarkably few known
examples of Seyfert nuclei in very late-type, low-mass disk galaxies.
The small number of NGC 4395 analogs in SDSS is largely a selection
effect: galaxies of this low luminosity are only sampled in the SDSS
spectroscopic survey out to $z\approx0.02-0.03$ due to the apparent
magnitude limit of $r=17.77$ of the spectroscopic survey.  Since SDSS
probes only a very small volume for these NGC 4395-type objects, our
sample is biased toward somewhat more luminous host galaxies and more
luminous AGNs.

Some limited information on stellar populations and star formation
history can be obtained from the \dn\ and \hdeltaa\ indices in the
MPA/JHU catalog.  The measurement of these parameters is described in
detail by \citet{kau03sfr}.  The \dn\ index measures the strength of
the 4000 \AA\ break, which increases monotonically with population age
for a single-burst population.  The \hdeltaa\ index is a measure of
the H$\delta$ absorption-line strength, which, for a single-burst
population, is strongest for ages of $\sim0.1-1$ Gyr.
\citet{kau03sfr} showed that population synthesis models for galaxies
incorporating exponentially declining star formation rates and random
bursts of star formation were generally able to track the locus of
SDSS galaxies in a plot of \hdeltaa\ vs.\ \dn.  To illustrate the
range of stellar population ages in our sample, Figure \ref{hostage}
displays the values of \hdeltaa\ and \dn\ from the MPA/JHU catalogs
for our sample, in comparison with the overall sample of all galaxies
in the catalog having $\log(\mstar/\msun)<10$.  The AGN hosts exhibit
a wide range in both stellar indices, but avoid the largest values of
\dn\ (corresponding to the oldest stellar populations).  In this
respect, they follow the same general trends as the overall population
of Seyfert 2 galaxies in SDSS, as shown by \citet{kew06}: the Seyfert
2s tend to be found in hosts with $\dn$ in the range $\sim1.2-1.8$,
while LINERs are found in systematically older host galaxies with \dn\
up to $\sim2.2$ \citep{kew06}.

\subsection{Black Hole Mass and Eddington Ratio}

For Seyfert 2 galaxies, there is no direct way to measure the black
hole mass from the optical spectrum of the AGN, and the best available
alternative is to use the stellar velocity dispersion and the \msigma\
relationship to obtain a rough estimate of \mbh.  As shown by
\citet{bgh05}, the low-mass Seyfert 1 galaxies from GH04 lie fairly
close to the extrapolated \msigma\ relation, on average having
slightly higher black hole masses than would be predicted from their
velocity dispersions.  The mean offset between the estimated black
hole masses for these galaxies and the \citet{tre02} \msigma\ relation
is 0.23 dex.  This offset is likely to reflect a flattening in the
slope of the \msigma\ relation at low masses \citep{w06,gh06}, but,
lacking a more detailed direct measurement of the change in the
\msigma\ slope as a function of \sigmastar, we simply assume a uniform
offset.  Since the Seyfert 2 galaxies in this sample fall in the same
range of stellar velocity dispersions as the GH04 Seyfert 1 galaxies,
we use the \sigmastar\ measurements to estimate \mbh, assuming an
offset of 0.23 dex relative to the Tremaine \etal\ \msigma\ relation.
For the objects not having direct measurements of \sigmastar, we use
\fwhmo/2.35 in place of \sigmastar\ to estimate \mbh.  With this
prescription, the black hole mass estimates for this sample are in the
range $4.7 < \log(\mbh/\msun) < 6.8$, with a median value of
$\log(\mbh/\msun) = 6.1$.  The uncertainty in any individual mass
estimate is likely to be at least 0.3 dex, however, due to the scatter
in the \msigma\ relation and the poorly constrained \msigma\ slope at
low masses.

Similarly, the bolometric luminosities can only be estimated
indirectly.  The best available indicator of the AGN luminosity in
these objects is the [\ion{O}{3}] $\lambda5007$ luminosity, which we
assume to arise predominantly from the AGN rather than from
star-forming regions, based on the locations of these galaxies in the
BPT diagnostic diagrams.  (As a caveat, we note that for host galaxies
of relatively low metallicity, the contribution from star-forming
regions to the [\ion{O}{3}] luminosity could be greater than in
high-metallicity galaxies, but we have no direct diagnostics of the
star-formation rate in the nuclei.)  From a sample of low-redshift
Seyfert 1 galaxies and quasars, \citet{hec04} found an [\ion{O}{3}]
bolometric correction of $\lbol/L_\mathrm{[OIII]} \approx 3500$, with
a typical scatter of 0.38 dex.  We adopt this prescription for our
sample, but note as an additional caveat that there is evidence that
the $L_\mathrm{[OIII]}/L_X$ ratio decreases systematically with AGN
luminosity \citep{net06}, implying that there is a luminosity
dependence to the [\ion{O}{3}] bolometric correction.  The
[\ion{O}{3}] luminosities for our sample are listed in Table
\ref{sample}; most are in the range $6.0 < \log(\lothree/\lsun) <
7.5$, implying bolometric luminosities of $43 < \log[\lbol/$(ergs
s\per)]$<44.6$ for most objects.  The two very low-redshift galaxies
with late-type spiral hosts, 1032+6502 and 1109+6123, are much lower
in luminosity, both having $\log(\lothree/\lsun) \approx 4.9$ and
$\lbol \approx 10^{42}$ ergs s\per.

Combining the black hole masses and luminosities, we can obtain rough
estimates of \lbol/\ledd.  The distribution of \lbol/\ledd\ estimates
is shown in Figure \ref{lbol}; the median value of $\log(\lbol/\ledd)$
for the sample is $-0.4$.  Although these values of \lbol/\ledd\ will
be extremely uncertain for individual galaxies, the estimates suggest
that many of these galaxies contain black holes that are undergoing a
significant episode of accretion.  \citet{gh07} found that in their
DR4 sample of Seyfert 1 galaxies with $\mbh < 2\times10^6$ \msun, the
distribution of Eddington ratio is peaked at $\log(\lbol/\ledd)
\approx -0.5$, highlighting the similarities between the two samples.

Alternatively, assuming the same [\ion{O}{3}] bolometric correction,
we can estimate the minimum black hole mass that would be required to
support the luminosity in each object if all objects are radiating at
$\lbol/\ledd<1$.  Figure \ref{bhmin} shows the resulting lower limits
to \mbh\ as a function of \sigmastar\ (again using \fwhmo/2.35 in
place of \sigmastar\ for the two objects without measured velocity
dispersions).  The minimum required black hole masses for the sample
are generally low, and 2/3 of the galaxies have \mbh(min)$<10^6$
\msun.  The results are consistent with most objects having black hole
masses that fall close to or below the low-mass extrapolation of the
\msigma\ relation from \citet{tre02}; eight galaxies in the sample
have minimum \mbh\ values that lie above the \msigma\ relation.  The
largest deviation above the extrapolated \msigma\ relation is found
for 1440+0247, which lies approximately an order of magnitude higher
than the Tremaine \etal\ \msigma\ relation extrapolated to
$\sigmastar=45$ \kms.  Given the substantial uncertainties in the
[\ion{O}{3}] bolometric correction, these lower limits to \mbh\ should
be viewed cautiously on a case-by-case basis, but on the whole these
results are consistent with expectations for a sample of black holes
of relatively low mass in which many of the black holes are undergoing
a significant accretion episode.

\subsection{Unified Models and the Obscuring Torus}

These Seyfert 2 galaxies and the GH04 Seyfert 1 objects do not
constitute perfectly matched samples in terms of their nuclear or host
galaxy properties, as a result of the very different selection
criteria that were used for the two samples.  Nevertheless, the two
samples do occupy similar and overlapping ranges in some key
parameters.  As shown in Figure \ref{hostcompare}, the distributions
of $g$-band absolute magnitudes are overlapping but not identical; the
Seyfert 2s were selected based on an absolute magnitude cut and all
have $M_g \geq -19.7$ mag, while the GH04 objects were selected
without regard to host galaxy properties, and their distribution
extends to slightly more luminous galaxies, up to $M_g = -20.6$.  The
GH04 sample also does not contain any host galaxies as faint as the
least luminous Seyfert 2 galaxy in this sample, 1109+6123, with $M_g =
-16.8$.  In terms of [\ion{O}{3}] luminosity, which is the best
available indicator of total AGN luminosity, the two samples occupy
overlapping ranges with most galaxies in the range $\lothree \sim
10^6-10^8$ \lsun; again, the Type 1 sample extends to slightly higher
luminosities and the Type 2 sample extends to lower luminosities, with
the galaxy 1109+6123 falling at the low-luminosity end of the
distribution.

Although the narrow-line ratios of the two samples occupy overlapping
ranges, as seen in Figure \ref{bptdiagrams}, there are some key
differences between the Type 1 and Type 2 objects.  From inspection of
Figure \ref{bptdiagrams}, it is evident that the Seyfert 2 galaxies
from this sample have somewhat lower values of [\ion{O}{3}]/\hbeta\ on
average than the GH04 sample.  the median values of
[\ion{O}{3}]/\hbeta\ are 7.9 and 4.9 for the Type 1 and Type 2 samples
respectively, and a Kolmogorov-Smirnov (K-S) test confirms that the
two distributions are significantly different, with a 1.4\%
probability of being drawn from the same parent population. We also
examine the [\ion{O}{2}]$\lambda3727$/[\ion{O}{3}]$\lambda5007$ ratio
for the two samples, measured from the SDSS spectra since the ESI
spectra do not cover [\ion{O}{2}] $\lambda3727$ for most objects (see
Figure \ref{hostcompare}).  The Type 1 and Type 2 objects have median
values for [\ion{O}{2}]/[\ion{O}{3}] of 0.23 and 0.61, respectively.
A K-S test shows that the two distributions of
[\ion{O}{2}]/[\ion{O}{3}] have only a $3\times10^{-5}$ probability of
being drawn from the same parent population.

The higher [\ion{O}{2}]/[\ion{O}{3}] ratio in the Type 2 objects is
intriguing in light of similar trends found recently among
high-luminosity AGNs.  \citet{ho05} showed that among the PG quasar
sample, the [\ion{O}{2}]/[\ion{O}{3}] ratio is generally consistent
with that expected from AGN photoionization, with no evidence for an
additional contribution to the [\ion{O}{2}] emission from star-forming
regions.  Following up on this study, \citet{khi06} found that
luminous Type 2 AGNs from SDSS such as those found by \citet{zak03}
have enhanced levels of [\ion{O}{2}] emission relative to what is seen
in a corresponding Type 1 AGN sample from SDSS, possibly indicating an
elevated rate of star formation in the Type 2 objects compared to the
Type 1 objects.  This observation presents a strong challenge to the
simplest unified models, in which Type 1 and Type 2 AGNs differ only
by orientation and obscuration of the central engine.  The low-mass
Seyfert 2 hosts evidently follow the same trend of enhanced
[\ion{O}{2}] emission: the median values of log([\ion{O}{3}]/\hbeta)
and log([\ion{O}{2}]/[\ion{O}{3}]) for the Type 2 objects in this
sample are 0.69 and $-0.21$, respectively, which fall remarkably close
to the values for these line ratios measured from the composite Type 2
quasar spectrum of \citet{zak03}, as displayed in Figure 7 of
\citet{khi06}.  The lower [\ion{O}{3}]/\hbeta\ ratios among the
Seyfert 2 sample could be another manifestation of a larger
contribution of star-forming regions to the spectra, although another
possibility is that the Seyfert 2 galaxies have slightly lower NLR
ionization parameter than the GH04 Seyfert 1 galaxies.

Aside from the question of the relative amounts of star formation in
the Type 1 and Type 2 samples, the more fundamental unification
question remains: do they all have intrinsically similar central
engines, in which case the presence or absence of broad emission lines
would depend primarily on their orientation and the degree of
obscuration along our line of sight?  Or, alternatively, are some of
the Type 2 objects ``true'' Type 2 AGNs that intrinsically lack a BLR?
There is growing observational evidence that at low luminosities,
there is a population of Seyfert 2 galaxies that do not exhibit
polarized broad-line emission or X-ray absorption
\citep[e.g.][]{pap01, tran03, nmm03, bian08}, suggesting that the line
of sight to the nucleus is unobscured and the BLR is not present.  One
scenario to explain true Seyfert 2 nuclei has been proposed by
\citet{nic00}; in this model, the BLR consists of an accretion disk
wind and when \lbol/\ledd\ falls below a critical value of
$\sim10^{-3}$, the luminosity is insufficient for the disk to form a
radiation-pressure dominated region from which to launch the wind.  In
another model proposed by \citet{laor03}, there is a maximum BLR
linewidth of $\sim25,000$ \kms\ above which the BLR clouds do not
survive, and there is a critical luminosity below which the BLR radius
shrinks to a size where the corresponding Keplerian velocities exceed
this maximum linewidth.  At present it is not completely clear whether
\lbol\ or \lbol/\ledd\ is the primary parameter that determines
whether a BLR forms (or whether there may be some additional key
parameter), and further exploration of the properties of the least
luminous Seyfert 2 galaxies is needed in order to resolve this
question.  As described above, most of our sample consists of objects
with relatively high \lbol/\ledd, and with \lbol\ in the range
$10^{43}-10^{44.6}$ ergs s\per; in both the \citet{nic00} and
\citet{laor03} scenarios these objects would be above the critical
luminosity or Eddington ratio needed to support BLRs.

A closely related question is the nature of the obscuring torus in
unified models of AGNs and whether the properties or existence of the
torus depends on AGN luminosity or other properties.  One attractive
possibility for the origin of the obscuring torus is that it may be
the dusty component of a wind driven from the surface of the accretion
disk \citep[e.g.,][]{kk94}.  \citet{es06} have recently examined the
dynamics of hydromagnetic disk winds and the dependence of outflow
properties on the AGN luminosity.  A key prediction of their model is
that for AGN luminosities below $\sim10^{42}$ ergs s\per, the AGN is
not powerful enough to sustain the torus wind, and the torus should
therefore not be present in very low-luminosity AGNs.  While most of
our sample is more luminous than this threshold, the least luminous
object in our sample (1109+6123) presents a particularly interesting
test case.  Its narrow-line spectrum and host galaxy are nearly
identical to NGC 4395 (\S\ref{individualobjects}), and it is the best
Type 2 counterpart to NGC 4395 known.  NGC 4395 has a bolometric
luminosity of $\sim5\times10^{40}$ ergs s\per\ \citep{mor99, pet05},
more than an order of magnitude below the typical cutoff luminosity
for a wind-driven torus in the Elitzur \& Shlosman model.  While the
bolometric luminosity of SDSS 1109+6123 is unknown, its [\ion{O}{3}]
luminosity is about 40\% smaller than that of NGC 4395.

This galaxy and NGC 4395 together pose an interesting challenge to
models for the obscuring torus: \emph{the Type 1/Type 2 dichotomy
apparently extends down to the very lowest luminosities seen among
nearby Seyfert galaxies}, below $\lbol=10^{41}$ ergs s\per.  While
this is already well established for massive host galaxies
\citep{ho08}, the discovery of a near-perfect Type 2 analog to NGC
4395 raises the possibility that the obscuring torus might still be
present even at these very low masses and luminosities.  The detection
of continuum polarization in 1109+6123 gives an intriguing hint of an
obscured nucleus, and further searches for the presence of an
obscuring torus and obscured BLR in 1109+6123, including deeper
spectropolarimetry, X-ray observations, and mid-infrared spectroscopy,
may provide critical new tests for wind-driven torus models.
Alternatively, 1109+6123 might be a ``true'' Seyfert 2 that
intrinsically lacks a BLR.  In this situation, the question remains:
why would a BLR occur in NGC 4395, but not in 1109+6123, which has a
similar [\ion{O}{3}] luminosity, and presumably a similar black hole
mass as well?  If 1109+6123 is a true unobscured Seyfert 2, this would
imply that the formation or lack of a BLR must depend on some other
properties of the AGN in addition to \lbol\ or \lbol/\ledd.

Obscuration by larger-scale dust lanes within the host galaxies is
likely to play some role as well in determining the relative numbers
of Type 1 and Type 2 AGNs in low-mass disk galaxies.  The SDSS images
show that three objects in our sample are nearly edge-on disk
galaxies, in which host galaxy-scale obscuration could easily explain
the lack of broad emission lines in the optical spectrum.  In
contrast, in the GH04 Seyfert 1 sample, none of the spiral host
galaxies are close to edge-on \citep{ghb08}.  This is consistent with
trends seen in the general Seyfert population; Seyfert 1 nuclei are
deficient in edge-on host galaxies relative to face-on hosts
\citep{keel80,mr95}.  Similarly, in deep X-ray surveys, the difference
between ``optically dull'' X-ray sources and broad-lined AGNs can be
attributed at least in part to host galaxy-scale obscuration related
to host inclination \citep{rig06}.  A search for nuclear dust lanes in
high-resolution images would provide some constraints on the impact of
host-galaxy scale obscuring material \citep[e.g.,][]{mgt98}.

\subsection{Notes on Individual Objects}
\label{individualobjects}

\emph{0110+0026}: This galaxy was selected from the MPA/JHU SDSS
catalog of narrow-line AGNs, but it has a very prominent broad \hal\
emission line visible in both the SDSS and ESI spectra.

\emph{0914+0238}: A possible companion is located about 25\arcsec\ to
the east of this galaxy, but its redshift is unknown (not catalogued
in NED).

\emph{0947+5349}: This galaxy appears to be part of an interacting
triplet, labeled in NED as CGCG 265--039.  The redshifts of the two
larger galaxies just south of it in the SDSS image are unknown.

\emph{1032+6502}: Also known as NGC 3259, this is a nearby SBbc
galaxy.  It was previously identified as a ROSAT source by
\citet{bol92} and \citet{mhh96}, and as a Seyfert by \citet{hao05a} from
the SDSS spectrum.  Our Keck spectrum confirms the presence of a weak
broad \hal\ emission line.  Multicolor \hst\ images show that it has a
substantial bulge, with $M \approx 3\times10^8$ \msun\ and a
predominantly old stellar population \citep{car07}.  It was also
recently noted as an example of a galaxy containing both a nuclear
star cluster and an AGN by \citet{seth08}.

\emph{1109+6123}: Also known as UGC 06192 or MCG +10-16-069, this
galaxy was not previously identified as hosting an AGN prior to SDSS.
It has an unusual high-excitation narrow-line spectrum with the lowest
[\ion{N}{2}]/\hal\ ratio among our sample ([\ion{N}{2}]/\hal$=0.18$),
characteristic of low-metallicity AGNs and very similar to the
spectrum of NGC 4395 \citep{kra99}, but without any hint of broad
emission lines.  Its [\ion{O}{3}] luminosity, from the MPA/JHU
catalog, is $\log(L_\mathrm{[OIII]}/\lsun) = 4.94$.  This is very
similar to the narrow-line luminosity of NGC 4395, which has
$\log(L_\mathrm{[OIII]}/\lsun) = 5.09$ based on the emission-line data
from \citet{kra99} and a distance of 4.3 Mpc \citep{thim04}.  With
$\log(\mstar/\msun) = 8.07$, this galaxy has the lowest stellar mass
of any unambiguous Seyfert 2 galaxy in either the DR2 or DR4 MPA/JHU
AGN catalogs.  From examination of the SDSS images, the host galaxy is
a low surface brightness, late-type spiral, probably best classified
as Scd or Sd.  A surface brightness profile measured from the SDSS
$i$-band image reveals that the host galaxy is well described by an
exponential disk with a scale length of 7\farcs9 ($\approx 1.1$ kpc),
with a slight central excess over the exponential profile in the inner
$\sim5\arcsec$ that may indicate the presence of a small bulge or
pseudobulge.  Overall, the properties of this galaxy make it a
near-twin of NGC 4395, aside from the lack of broad emission lines in
its spectrum.  Deeper and higher-resolution imaging with \hst\ would
be particularly valuable to definitively detect or constrain the
presence of a bulge in this galaxy and to determine whether a compact
central star cluster is present.

\emph{1432+0046}: The SDSS image shows an apparent close companion
just to the east of this galaxy, but it is a chance projection; the
disk galaxy is SDSS J143232.24+004617.4, at $z=0.034$.

\emph{1440+0247}: Also known as Tol 1437+030, this object was
identified as an AGN by \citet{bfw78}.  It has also been noted as an
AGN from the SDSS spectrum by \citet{and03}, \citet{kni04}, and
\citet{hao05a}.  Its spectrum is very similar to those of NGC 4395 and
POX 52 in terms of the narrow-line ratios as well as the detection of
high-ionization lines (up to [\ion{Fe}{10}] and [\ion{Fe}{11}]), but
broad \hal\ emission is only marginally detectable in the ESI
spectrum.

\emph{1629+4254} : At first glance this appears to be a likely Seyfert
1 galaxy because of the broad base to the \hal+[\ion{N}{2}] emission
blend, but our line profile fits show that the broad bases are present
on all of the forbidden lines.  The [\ion{O}{3}] profile has an
extraordinary broad and redshifted component (FWHM = 589 \kms, $\Delta
v = +223$ \kms\ relative to the core component), while the core
component is fairly narrow (FWHM = $78\pm1$ \kms).  From the FIRST
radio detection, the radio power of this source at 20 cm is
$5\times10^{21}$ W Hz\per.  Multicomponent [\ion{O}{3}] profiles are
often found in objects with linear radio sources \citep{whi88}, and
the detection of a radio counterpart in the FIRST survey makes this an
unusual object relative to the very radio-quiet GH04 sample.  To
compare with the radio properties of the GH04 objects, we follow
\citet{ghu06} and adopt the radio loudness parameter $R =
f_\mathrm{6~cm}/f_\mathrm{4400~\AA}$.  We estimate the flux density of
the (presumably obscured) optical nonstellar continuum by using the
[\ion{O}{3}] flux as a proxy, assuming the median ratio of
[\ion{O}{3}] luminosity to 4400 \AA\ continuum luminosity for the GH04
sample \citep{ghu06}.  This yields $f$(4400 \AA) $\approx 0.6$ mJy.
Based on the FIRST 20 cm flux and an assumed flat spectral index
between 6 and 20 cm, which is typical of low-luminosity Seyfert 2
galaxies \citep{uh01}, we find $R \approx 3$.  This is formally in the
radio-quiet regime (since $R<10$), but this value of $R$ is greater
than most of the upper limits found for the GH04 sample \citep{ghu06}.

\section{Summary and Conclusions}

We have presented an initial, exploratory study of the properties of
Seyfert 2 galaxies with sub-\lstar\ luminosities selected from the
Sloan Digital Sky Survey.  Our measurements reveal very low central
stellar velocity dispersions (\sigmastar $<60$ \kms) in 12 objects;
these are among the smallest velocity dispersions found in any AGN
host galaxies, and imply that these galaxies contain some of the least
massive black holes in any known AGNs.  We have also identified one of
the very few known examples of a high-excitation Seyfert 2 nucleus in
a late-type, dwarf spiral host galaxy similar to NGC 4395.  The
correlations between [\ion{O}{3}] linewidth and \sigmastar\
established for higher-mass Seyferts continue to hold in this low-mass
regime, while the low-mass Seyfert 2 galaxies exhibit an unusually
high incidence of peculiarities (such as redward asymmetries and
double-peaked profiles) in their forbidden emission lines.  Using
rough estimates of black hole mass and bolometric luminosity, we find
that the median value of $\lbol/\ledd$ for this sample is 0.4,
indicating that many of these are objects in which the black hole is
undergoing a major growth phase.  Future work on these objects will
include \emph{XMM-Newton} observations and \emph{Spitzer} mid-infrared
spectroscopy, in order to search for evidence of nuclear obscuration
and to test whether the obscuring torus of AGN unified models is
present in this low-mass, low-luminosity regime.

\acknowledgments 

We are very grateful to Guinevere Kauffmann and her collaborators for
making their AGN and stellar mass catalogs available to the community.
We thank George Djorgovski for obtaining some preliminary ESI spectra
for this project, Craig Markwardt and Mike Blanton for their excellent
and useful IDL software, Michael Strauss for providing plotting
scripts, and an anonymous referee for suggestions that improved the
manuscript.  Research by A.J.B. was supported by the UC Irvine
Physical Sciences Innovation Fund and by NSF grant AST-0548198.  Data
presented herein were obtained at the W.M. Keck Observatory, which is
operated as a scientific partnership among Caltech, the University of
California, and NASA. The Observatory was made possible by the
generous financial support of the W.M. Keck Foundation.  The authors
wish to recognize and acknowledge the very significant cultural role
and reverence that the summit of Mauna Kea has always had within the
indigenous Hawaiian community. Funding for the SDSS and SDSS-II has
been provided by the Alfred P. Sloan Foundation, the Participating
Institutions, the National Science Foundation, the U.S. Department of
Energy, the National Aeronautics and Space Administration, the
Japanese Monbukagakusho, the Max Planck Society, and the Higher
Education Funding Council for England. The SDSS Web Site is
http://www.sdss.org/.


\clearpage

\begin{deluxetable}{lccccccccc}
\tablewidth{7in} \tablecaption{Sample Properties}

\tablehead{ \colhead{Galaxy} & \colhead{Plate/MJD/Fiber} & \colhead{z} &
     \colhead{$M_g$} & \colhead{$\log{M_\star}$} & \colhead{log $L$([O III])} &
     \colhead{$C$} & \colhead{References}  \\
     \colhead{} & \colhead{} &  \colhead{} & \colhead{(mag)} &
     \colhead{(\msun)} &
     \colhead{($L_\odot$)} & & }

\startdata
\object{SDSS J010053.57+152728.1} & 421.51821.461 & 0.0404 & $-19.5$ & 9.75 & 5.83 & 2.59 & a,b \\
\object{SDSS J011059.31+002601.1} & 397.51794.474 & 0.0188 & $-18.7$ & 9.81 & 6.23 & 2.46 & a \\
\object{SDSS J011905.14+003745.0} & 398.51789.585 & 0.0327 & $-17.6$ & 9.16 & 6.64 & 3.26 & a,b \\
\object{SDSS J021405.91-001637.0} & 405.51816.168 & 0.0373 & $-18.9$ & 9.72 & 7.37 & 2.74 & a,b \\
\object{SDSS J034330.26-073507.4} & 462.51909.044 & 0.0357 & $-18.7$ & 9.61 & 5.81 & 2.84 & a,b \\
\object{SDSS J091414.34+023801.8} & 567.52252.259 & 0.0735 & $-19.7$ & 9.97 & 7.56 & 2.78 & a \\
\object{SDSS J091608.50+502126.9} & 766.52247.197 & 0.0495 & $-18.8$ & 9.90 & 6.39 & 2.64 & a \\
\object{SDSS J094716.13+534944.9} & 769.52282.240 & 0.0383 & $-18.5$ & 9.56 & 6.78 & 2.85 & a \\
\object{SDSS J100654.83+445642.8} & 943.52376.222 & 0.0419 & $-19.0$ & 10.1 & 6.99 & 3.20 & a \\
\object{SDSS J102332.00+645240.2} & 489.51930.308 & 0.0404 & $-19.2$ & 9.78 & 6.17 & 2.76 & a,b \\
\object{SDSS J103126.56+624648.6} & 772.52375.548 & 0.0529 & $-19.4$ & 10.50 & 7.55 & 2.49 & a \\
\object{SDSS J103234.85+650227.9} & 489.51930.193 & 0.0056 & $-18.6$ & \nodata & 4.85 & 2.29 & b,c  \\
\object{SDSS J105308.99+041036.2} & 579.52338.107 & 0.0431 & $-18.5$ & 9.54 & 6.61 & 2.61 & a \\
\object{SDSS J110912.40+612346.7} & 775.52295.309 & 0.0067 & $-16.8$ & 8.07 & 4.94 & 1.95 & a \\
\object{SDSS J111255.26+552928.8} & 909.52379.249 & 0.0498 & $-19.1$ & 9.86 & 6.53 & 2.53 & a \\
\object{SDSS J114315.36+631108.2} & 776.52319.604 & 0.0622 & $-19.4$ & 9.68 & 7.16 & 2.37 & a \\
\object{SDSS J120815.94+512325.7} & 882.52370.012 & 0.0331 & $-18.3$ & 9.60 & 6.54 & 2.45 & a \\
\object{SDSS J130525.75+642121.5} & 602.52072.172 & 0.0527 & $-19.5$ & 10.41 & 7.37 & 2.41 & a,b \\
\object{SDSS J142151.63+033121.2} & 584.52049.218 & 0.0552 & $-18.9$ & 9.59 & 6.66 & 2.76 & a \\
\object{SDSS J143231.15+004614.4} & 535.51999.049 & 0.0712 & $-19.5$ & 9.97 & 7.65 & 2.50 & a,b \\
\object{SDSS J143534.96+591658.3} & 790.52441.409 & 0.0682 & $-19.4$ & 10.41 & 7.51 & 2.48 & a \\
\object{SDSS J144012.70+024743.5} & 536.52024.575 & 0.0298 & $-19.0$ & \nodata & 7.43 & 2.72 & b,d,e  \\ 
\object{SDSS J151135.60+023807.0} & 540.51996.524 & 0.0389 & $-19.2$ & 9.71 & 6.89 & 2.83 & a,b \\
\object{SDSS J155404.40+545708.2} & 619.52056.448 & 0.0457 & $-18.8$ & 9.98 & 6.46 & 2.45 & a,b\\
\object{SDSS J160428.50-010435.7} & 344.51693.053 & 0.0296 & $-19.4$ & 9.78 & 6.45 & 2.74 & a \\
\object{SDSS J160510.73+472901.5} & 813.52354.558 & 0.0435 & $-19.0$ & 9.85 & 6.08 & 2.85 & a \\
\object{SDSS J161038.31+522334.2} & 623.52051.381 & 0.0617 & $-19.4$ & 10.04 & 6.51 & 1.95 &a,b  \\
\object{SDSS J162917.44+425439.8} & 815.52374.032 & 0.0353 & $-18.1$ & 9.61 & 7.44 & 2.24 & a \\
\object{SDSS J172344.17+583032.2} & 366.52017.274 & 0.0780 & $-19.7$ & 10.16 & 7.61 & 2.55 & a,b
\enddata
\tablecomments{``Plate/MJD/Fiber'' gives the reference to the SDSS
  spectroscopic observation of each object, for cross-identification
  in the SDSS archives.  References for previous AGN identifications
  are: (a) \citet{kau03agn} with updated data from
  http://www.mpa-garching.mpg.de/SDSS ; (b) \citet{hao05a}; (c)
  \citet{mhh96}; (d) \citet{bfw78}; (e) \citet{and03}.  Stellar
  masses, [\ion{O}{3}] luminosities, velocity dispersions, and
  concentration indices $C$ are from the Kauffmann \etal\ online
  catalog, except for 1032+6502 and 1440+0247; for these two objects
  the concentration indices are from the SDSS database and
  $L$([\ion{O}{3}]) is determined from the extinction-corrected
  [\ion{O}{3}] flux and the galaxy distance.  
  Absolute $g$-band magnitudes are corrected for Galactic
  extinction using the \citet{sfd98} extinction maps and $k$-corrected
  using M.\ Blanton's $kcorrect$ code.  }
\label{sample}
\end{deluxetable}

\clearpage

\begin{deluxetable}{lcccccc}
\tablewidth{7in} \tablecaption{Velocity Dispersions and Linewidths}

\tablehead{ \colhead{Galaxy} & \colhead{Exp. Time} & \colhead{S/N} &
  \colhead{$\sigmastar$} & \colhead{$\sigma$([O~III])} & 
     \colhead{FWHM([O~III])} & \colhead{FWHM([S~II])} \\ \colhead{} & \colhead{(s)} &
     \colhead{} & \colhead{(\kms)} & \colhead{(\kms)} & \colhead{(\kms)} & \colhead{(\kms)}  }

\startdata
J010053.57$+$152728.1 & 900  & 10.9 & $52\pm4$ &  $36\pm2$  &$86\pm4$  & $77\pm9$ \\
J011059.31$+$002601.1 & 3600 & 40.4 & $63\pm5$ &  $88\pm3$  &$143\pm11$ & $157\pm2$ \\
J011905.14$+$003745.0 & 900  & 8.4  & $39\pm8$ &  $66\pm2$  &$77\pm3$  & $90\pm3$ \\
J021405.91$-$001637.0 & 3600 & 35.6 & $59\pm5$ &  $181\pm1$ &$159\pm1$ & $142\pm2$ \\
J034330.26$-$073507.4 & 3600 & 45.8 & $94\pm3$ &  $75\pm1$  & $223\pm3$\tablenotemark{a} & \nodata\tablenotemark{a} \\
J091414.34$+$023801.8 & 1800 & 6.1  & $65\pm12$&  $68\pm1$  &$152\pm4$\tablenotemark{a} & $164\pm13$ \\
J091608.50$+$502126.9 & 1800 & 11.6 & $48\pm3$ &  $75\pm3$  &$96\pm8$ & $75\pm2$  \\
J094716.13$+$534944.9 & 1800 & 14.5 & $55\pm3$ &  $77\pm1$  &$147\pm1$ & $129\pm1$ \\
J100654.83$+$445642.8 & 1800 & 12.0 & $65\pm5$ &  $147\pm13$ &$428\pm7$\tablenotemark{a} & \nodata\tablenotemark{a} \\
J102332.00$+$645240.2 & 1800 & 16.8 & $56\pm5$ &  $57\pm1$  &$145\pm3$  & $161\pm5$ \\
J103126.56$+$624648.6 & 1800 & 7.1  & $72\pm9$ &  $85\pm4$  &$138\pm13$ & $236\pm5$  \\
J103234.85$+$650227.9 & 1800 & 25.2 & $43\pm4$ &  $69\pm1$  &$71\pm4$  & $53\pm2$ \\
J105308.99$+$041036.2 & 1800 & 2.6  & \nodata  &  $84\pm8$  &$134\pm24$ & $125\pm18$  \\
J110912.40$+$612346.7 & 1700 & 3.4  & \nodata  &  $28\pm1$  &$66\pm1$ & $63\pm2$ \\
J111255.26$+$552928.8 & 1800 & 12.7 & $67\pm4$ &  $69\pm1$  &$109\pm4$ & $128\pm5$  \\
J114315.36$+$631108.2 & 1800 & 14.7 & $70\pm9$ &  $123\pm2$ &$168\pm3$ & $152\pm8$ \\
J120815.94$+$512325.7 & 1800 & 4.6  & $47\pm8$ &  $74\pm2$  &$174\pm8$ & $153\pm7$ \\
J130525.75$+$642121.5 & 1800 & 9.9  & $77\pm4$ &  $70\pm3$  &$166\pm8$ & $218\pm4$ \\
J142151.63$+$033121.2 & 1800 & 18.4 & $50\pm4$ &  $62\pm1$  &$71\pm2$  & $66\pm2$ \\
J143231.15$+$004614.4 & 1800 & 7.9  & $65\pm8$ &  $75\pm6$  &$138\pm14$ & $118\pm9$ \\
J143534.96$+$591658.3 & 1800 & 11.1 & $91\pm6$ &  $96\pm4$  &$162\pm10$ & $172\pm7$ \\
J144012.70$+$024743.5 & 3600 & 32.0 & $45\pm4$ &  $122\pm1$ &$108\pm1$ & $105\pm1$ \\ 
J151135.60$+$023807.0 & 1800 & 29.3 & $65\pm5$ &  $95\pm2$  &$131\pm4$ & $135\pm6$ \\
J155404.40$+$545708.2 & 1800 & 25.0 & $73\pm3$ &  $62\pm1$  &$72\pm1$  & $63\pm3$ \\
J160428.50$-$010435.7 & 1800 & 20.2 & $56\pm4$ &  $97\pm2$  &$94\pm4$  & $103\pm2$ \\
J160510.73$+$472901.5 & 1800 & 19.7 & $67\pm3$ &  $149\pm5$ &$217\pm14$ & $197\pm6$ \\
J161038.31$+$522334.2 & 1800 & 8.1  & $59\pm7$ &  $88\pm4$  &$152\pm12$ & $158\pm9$  \\
J162917.44$+$425439.8 & 1800 & 17.3 & $61\pm4$ &  $238\pm1$ &$90\pm2$\tablenotemark{a} & $141\pm2$\tablenotemark{a,b} \\
J172344.17$+$583032.2 & 1800 & 15.0 & $66\pm6$ &  $105\pm3$ &$193\pm6$ & $154\pm3$
\enddata

\tablecomments{S/N is the mean signal-to-noise ratio per pixel in the
  extracted ESI spectrum over the wavelength range 5200--5600 \AA. }
  \tablenotetext{a}{[O III] profile is double-peaked.}
\label{esidata}
\end{deluxetable}

\clearpage

\begin{deluxetable}{lc}
\tablewidth{3in}
\tablecaption{Emission Lines in the Composite Seyfert 2 Spectrum}
\tablehead{ \colhead{Line} & \colhead{$f/f(\hbeta)$} }
\startdata
{}[\ion{Ne}{3}] $\lambda$3868   &     $0.55  \pm    0.04$ \\
{}[\ion{Ne}{3}] $\lambda$3967   &     $0.15  \pm    0.01$ \\
H$\epsilon$                      &     $0.15  \pm    0.01$ \\
{}[\ion{S}{2}] $\lambda$4068    &     $0.08  \pm    0.01$ \\
H$\delta$                        &     $0.22  \pm    0.02$ \\
H$\gamma$                        &     $0.41  \pm    0.02$ \\
{}[\ion{O}{3}]    $\lambda$4363 &     $0.07  \pm    0.01$ \\
\ion{He}{1}  $\lambda$4471       &     $0.03  \pm    0.01$ \\
\ion{He}{2} $\lambda$4686        &     $0.20  \pm    0.02$ \\
{}[\ion{Ar}{4}] $\lambda$4711   &     $0.04  \pm    0.01$ \\
{}[\ion{Ar}{4}] $\lambda$4740   &     $0.04  \pm    0.01$ \\
\hbeta\                          &     $1.00  \pm    0.03$ \\
{}[OIII] $\lambda$4959          &     $2.07  \pm    0.06$ \\
{}[OIII] $\lambda$5007          &     $6.47  \pm    0.16$ \\
{}[N I] $\lambda$5199,5201      &     $0.12  \pm    0.01$ \\
\ion{He}{1} $\lambda$5876        &     $0.12  \pm    0.01$ \\
{}[\ion{Fe}{7}] $\lambda$6086   &     $0.04  \pm    0.01$ \\
{}[\ion{O}{1}] $\lambda$6300    &     $0.52  \pm    0.02$ \\
{}[\ion{S}{3}] $\lambda$6312    &     $0.05  \pm    0.01$ \\
{}[\ion{O}{1}] $\lambda$6363    &     $0.22  \pm    0.02$ \\
{}[\ion{N}{2}] $\lambda$6548    &     $0.95  \pm    0.11$ \\
\hal\                            &     $4.63  \pm    0.15$ \\
{}[\ion{N}{2}] $\lambda$6583    &     $2.64  \pm    0.12$ \\
\ion{He}{1} $\lambda$6678        &     $0.02  \pm    0.01$ \\
{}[\ion{S}{2}] $\lambda$6716    &     $1.18  \pm    0.03$ \\
{}[\ion{S}{2}] $\lambda$6731    &     $0.99  \pm    0.03$ \\
{}[\ion{Ar}{3}] $\lambda$7135   &     $0.19  \pm    0.01$ \\
{}[\ion{O}{2}] $\lambda$7310    &     $0.09  \pm    0.01$ \\
{}[\ion{O}{2}] $\lambda$7331    &     $0.10  \pm    0.01$ \\
{}[\ion{S}{3}] $\lambda$9069    &     $0.74  \pm    0.05$ \\
{}[\ion{S}{3}] $\lambda$9532    &     $1.93  \pm    0.12$ \\
\enddata
\tablecomments{ Line strengths measured from the composite Seyfert 2
 spectrum are given relative to \hbeta.  The spectrum is corrected for
 Galactic extinction but not for internal extinction within the host
 galaxies.}
\label{compositelines}
\end{deluxetable}

\clearpage

\begin{deluxetable}{lccccccc}
\tablewidth{7in} \tablecaption{Polarization Data}

\tablehead{ \colhead{Galaxy} & \colhead{S/N} & \colhead{$p_b$} & \colhead{$\theta_b$}
    & \colhead{$p_r$} & \colhead{$\theta_r$} & \colhead{$E(B-V)$} &
    \colhead{$p_\mathrm{max}\mathrm{(Galactic)}$}
\\ \colhead{} & \colhead{} & 
    \colhead{(\%)} & \colhead{(degrees)} & \colhead{(\%)} &
    \colhead{(degrees)} & \colhead{(mag)} & \colhead{($\%$)}
}

\startdata
J011905.14$+$003745.0 & 80 & $0.55\pm0.06$ & $111\pm3$ & $0.29\pm0.05$ & $119\pm6$ & 0.032 & 0.29 \\
J021405.91$-$001637.0 & 160 & $0.50\pm0.03$ & $160\pm2$ & $0.52\pm0.03$ & $159\pm2$ & 0.038 & 0.34 \\
J094716.13$+$534944.9 & 105 & $0.22\pm0.04$ & $33\pm5$ & $0.20\pm0.04$ & $25\pm5$ & 0.014 & 0.13 \\
J110912.40$+$612346.7 & 42 & $0.76\pm0.10$ & $69\pm4$ & $0.72\pm0.10$ & $72\pm4$ & 0.011 & 0.10 \\
HD 57702 (null) & 1060 & $0.07\pm0.01$ & $161\pm1$ & $0.08\pm0.01$ & $150\pm2$ \\
\enddata 
\tablecomments{S/N is the signal-to-noise per pixel over the
continuum wavelength range 6380--6480 \AA\ in the total flux
spectrum. Polarization on the blue and red sides ($p_b$, $p_r$) and
polarization angles ($\theta_b$, $\theta_r$) are measured over the
wavelength regions 4000--5500 and 6000--7500 \AA.  Galactic reddening estimates
are from \citet{sfd98}, and the maximum estimated Galactic foreground
polarization is $p_\mathrm{max} (\%) = 9.0\times E(B-V)$
\citep{smf75}.
\label{poldata}}
\end{deluxetable}

\clearpage

\begin{figure}
\plotone{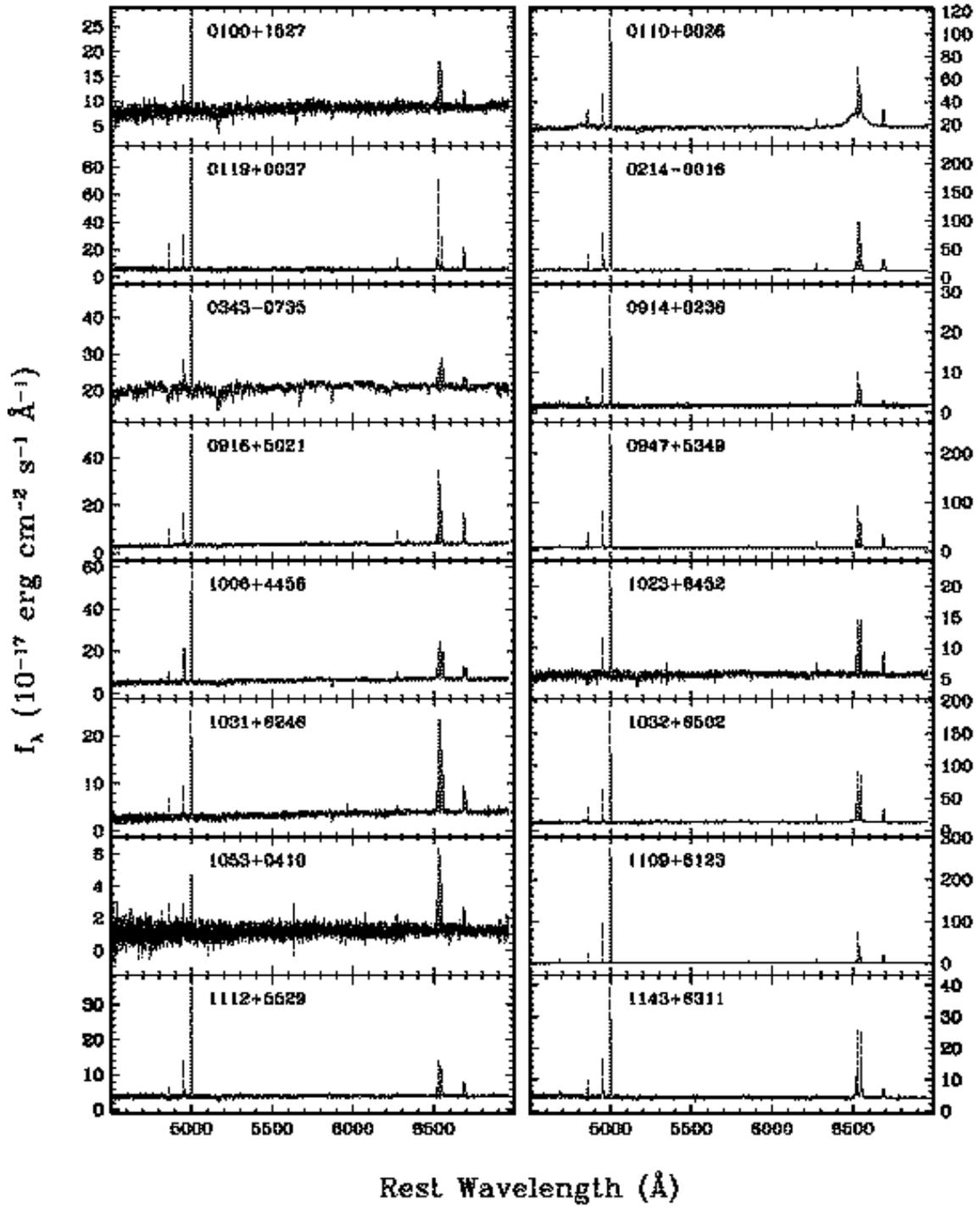}
\caption{Central portions of the Keck ESI spectra of the SDSS Seyfert
sample.  The spectra are binned to 0.5 \AA\ pixel\per\ for clarity.
\label{spectra}}
\end{figure}

\addtocounter{figure}{-1}
\begin{figure}
\plotone{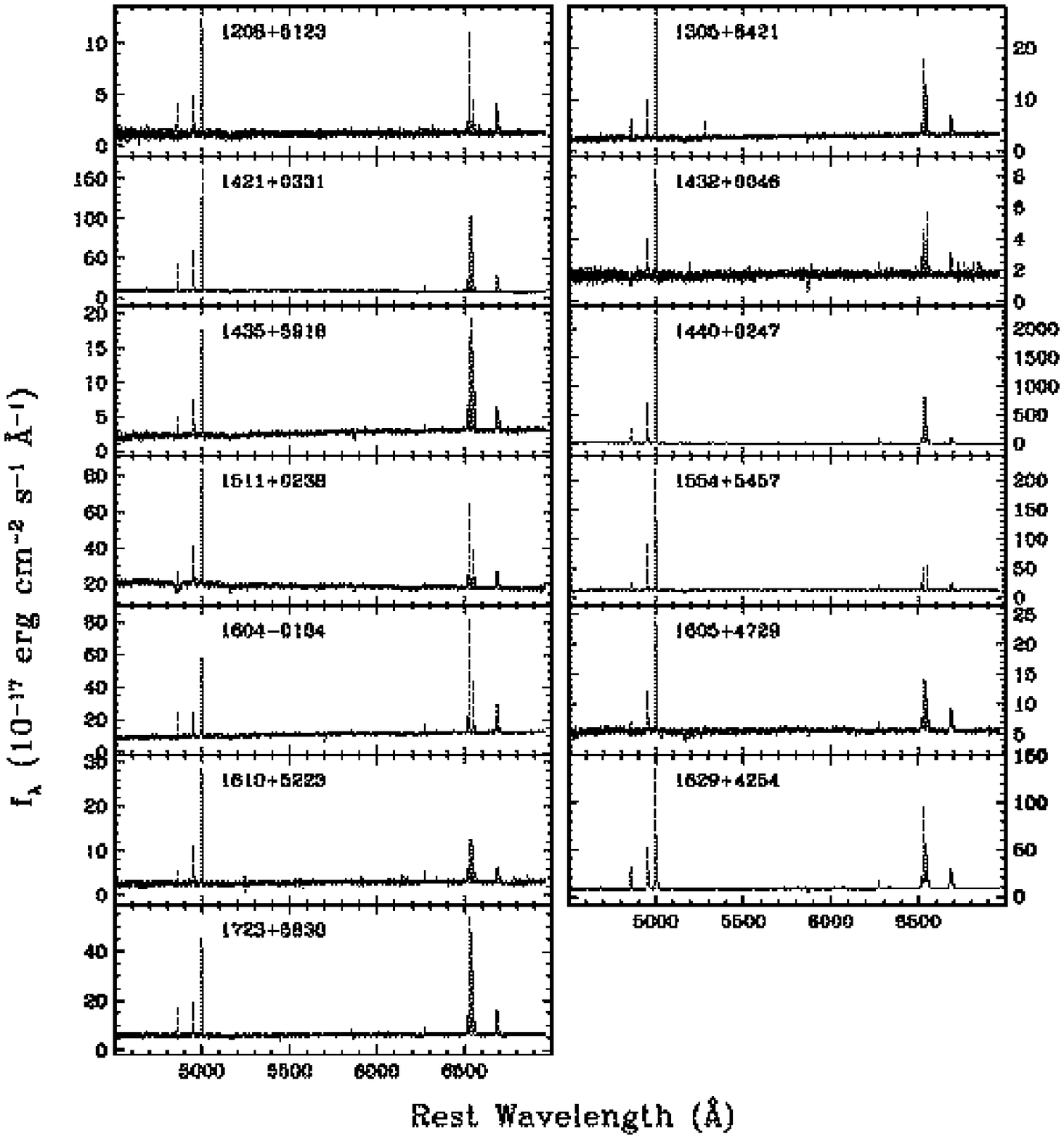}
\caption{(Continued) Keck ESI spectra of the SDSS Seyfert sample.}
\end{figure}

\begin{figure}
\plotone{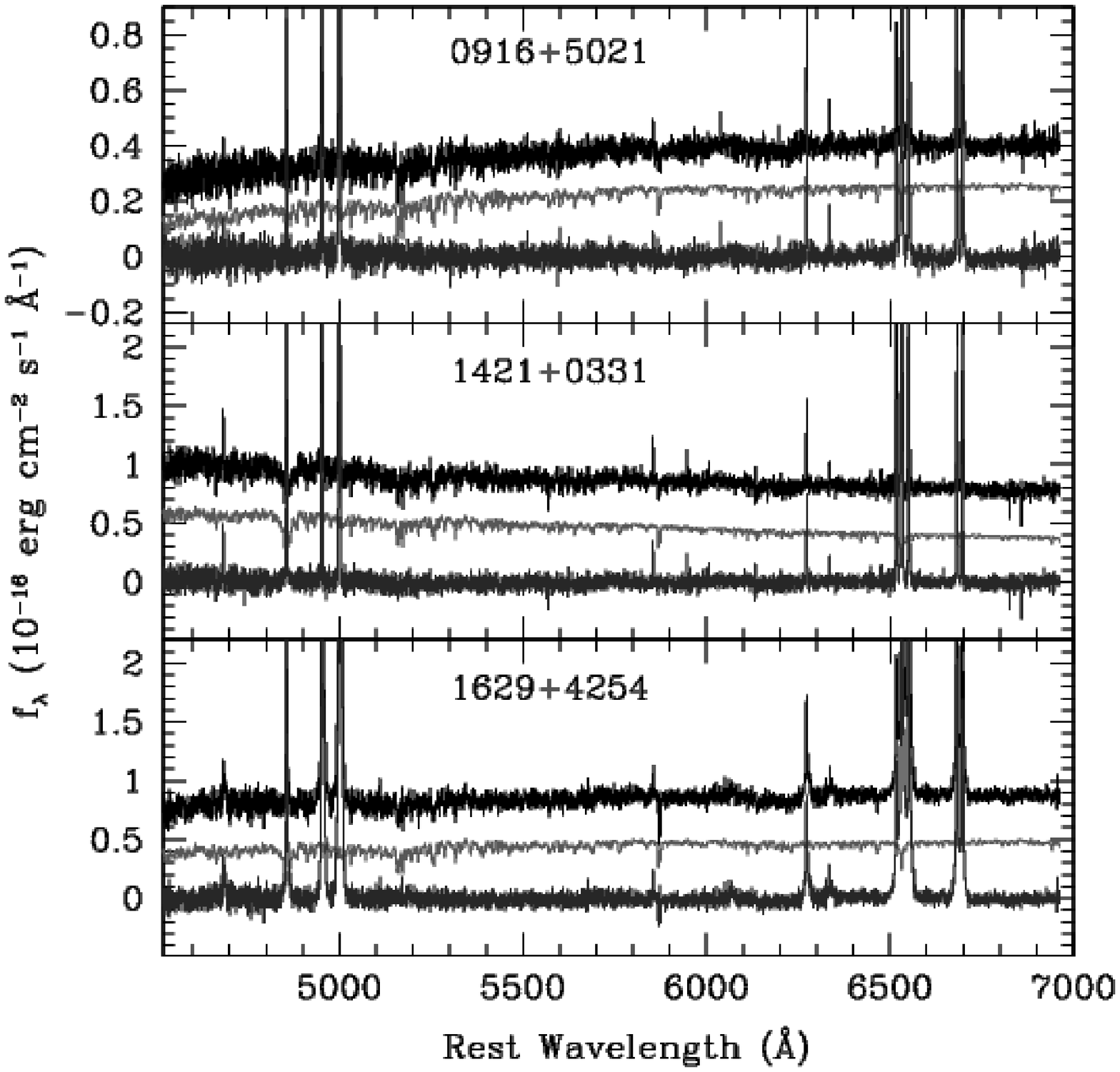}
\caption{Examples of starlight subtraction.  In each panel, the galaxy
  spectrum is in black, the best-fitting model is in red, and the
  starlight-subtracted spectrum is in blue.  The model spectra are
  shifted downward by a constant offset for clarity.
\label{starsub}}
\end{figure}

\begin{figure}
\plotone{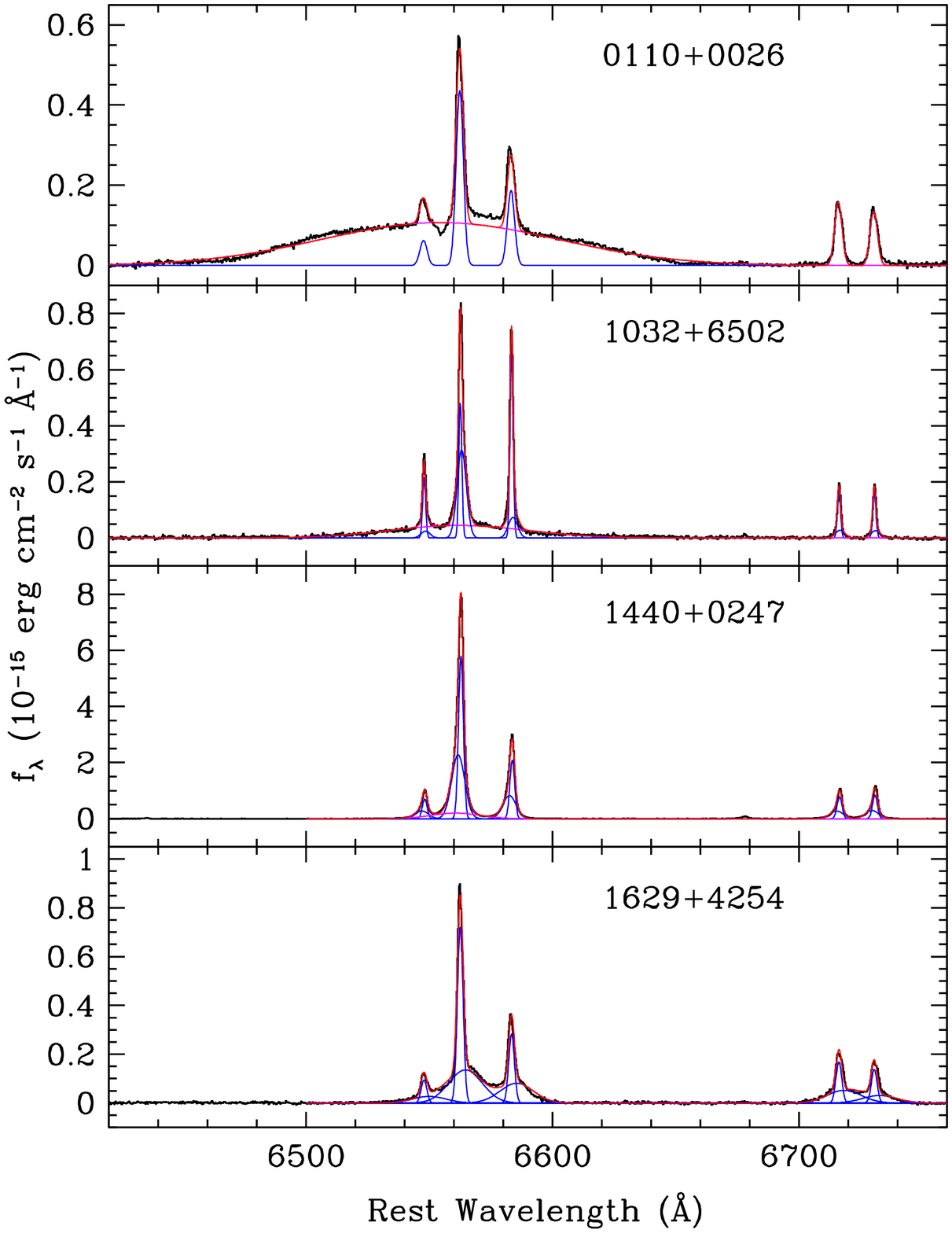}
\caption{Multi-Gaussian fits to the \hal, [\ion{N}{2}], and
  [\ion{S}{2}] lines for four objects, after starlight subtraction.
  In each plot, the narrow-line components are plotted in blue, the
  broad components in magenta, and the sum of all model components in
  red.  For the galaxy 1629+4254, all of the narrow lines contain the
  redshifted wing component, and the \hal\ profile does not require an
  additional broad component.
\label{broadhal}
}
\end{figure}

\begin{figure}
\begin{center}
\rotatebox{-90}{\scalebox{0.7}{\includegraphics{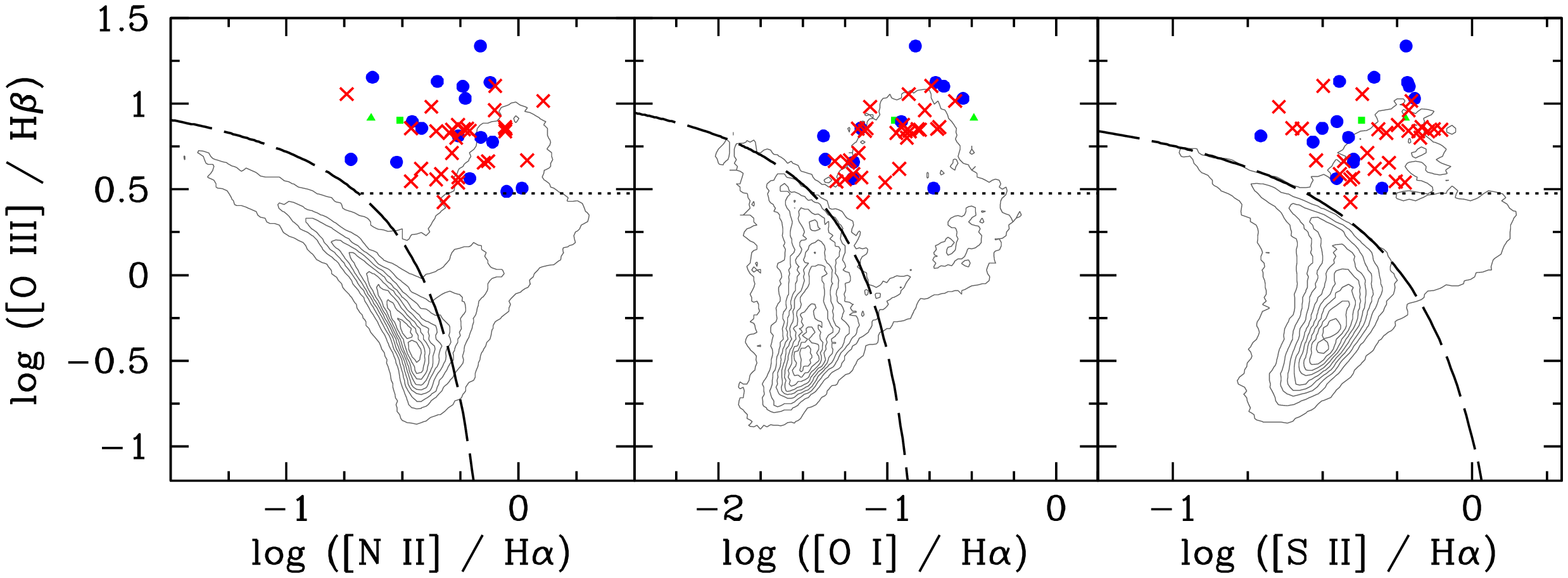}}}
\end{center}
\caption{Line-ratio diagrams for [O III] $\lambda5007$ / \hbeta\
  versus [N II] $\lambda6583$ / \hal, [S II] $\lambda6716,6731$ /
  \hal, and [O I] $\lambda6300$ / \hal.  Grey contours represent
  emission-line galaxies from \cite{kau03agn} with S/N $>6$ in
  \hal, [\ion{O}{3}], and [\ion{N}{2}], and S/N$>3$ in
  [\ion{O}{1}] flux.  Blue circles are the Seyfert 1 galaxies from
  \citet{gh04}, and red crosses are measurements from the Keck data
  for the Seyfert sample described here.  NGC 4395 and POX 52 are
  shown as a green triangle and square.  Dashed curves are the
  ``maximum starburst'' lines from \citet{kew06}, and the horizontal
  dotted line shows the selection cut at [\ion{O}{3}]/\hbeta\ $=3$.
\label{bptdiagrams}}
\end{figure}

\begin{figure}
\plotone{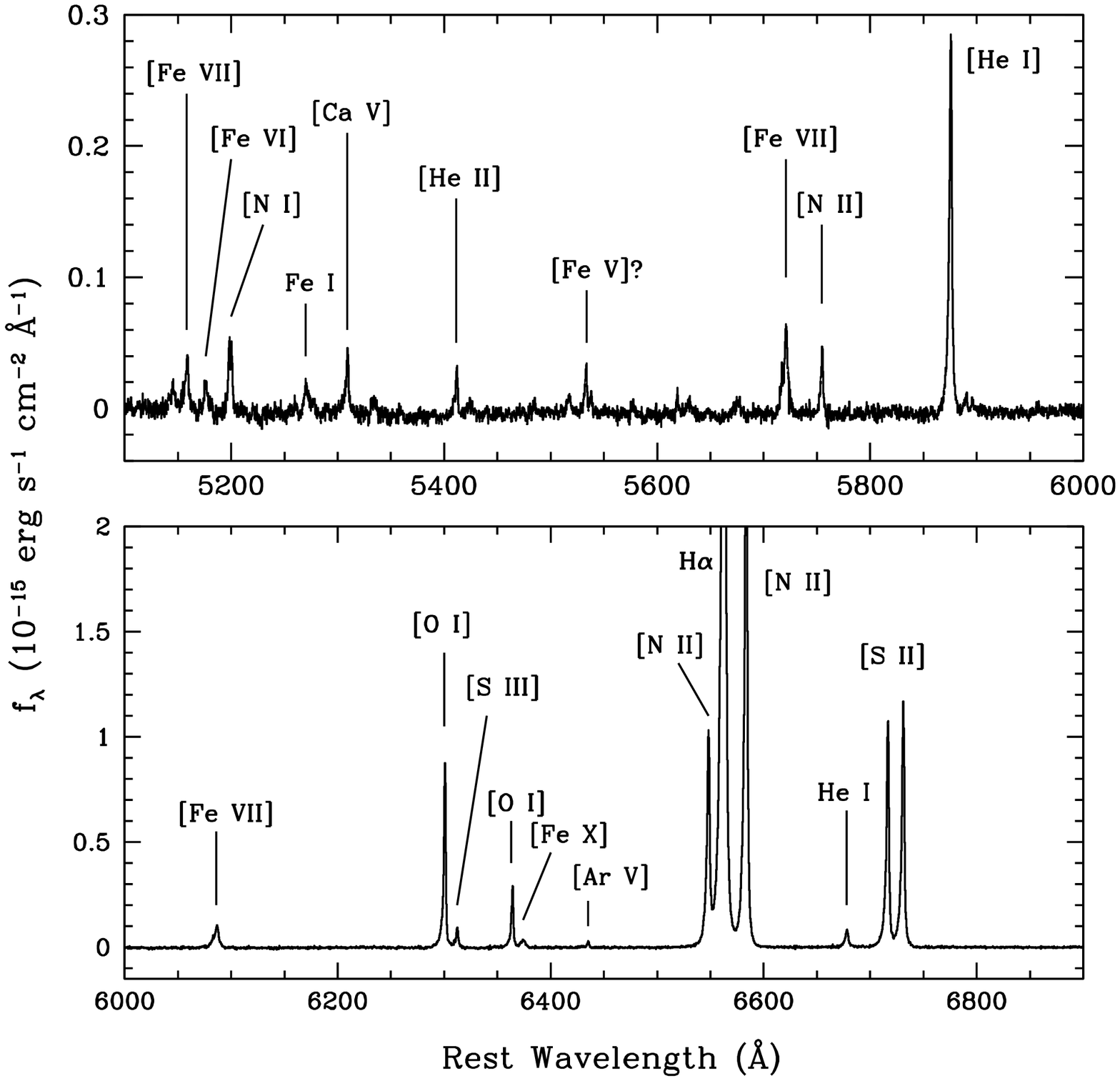}
\caption{Two sections of the starlight-subtracted ESI spectrum of SDSS
  J144012.70+024743.5, showing coronal-line emission including lines
  of [\ion{Fe}{7}] and [\ion{Fe}{10}].  [\ion{Fe}{11}] emission is
  also present at 7892 \AA.
\label{1440lines}
}
\end{figure}

\begin{figure}
\begin{center}
\scalebox{0.7}{\rotatebox{-90}{\includegraphics{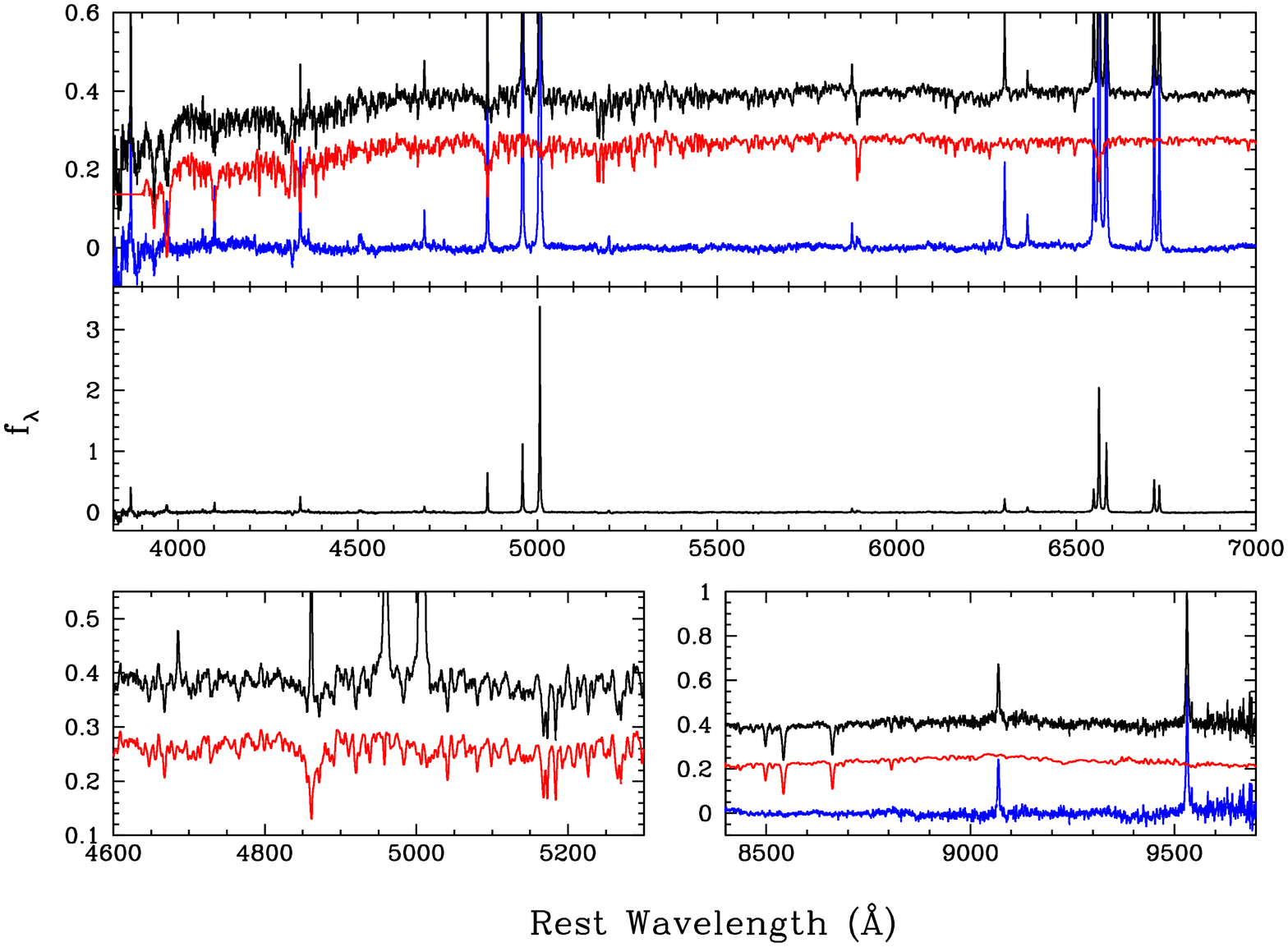}}}
\end{center}
\caption{Composite Seyfert 2 spectrum.  \emph{Upper panel:}  Composite
  spectrum (in black), model fit from the starlight subtraction
  routine (red), and starlight-subtracted spectrum (blue).
  \emph{Middle panel:}  Expanded view of the emission-line spectrum
  after starlight subtraction.  \emph{Bottom panels:}  Expanded views
  of the region including \hbeta\ and the Mg$b$ lines (left) and the
  \ion{Ca}{2} triplet and [\ion{S}{3}] $\lambda\lambda$9069, 9532 \AA\
  emission lines.  The model spectra (red) are shifted downward by a
  constant offset for clarity.
\label{composite}}
\end{figure}

\begin{figure}
\plotone{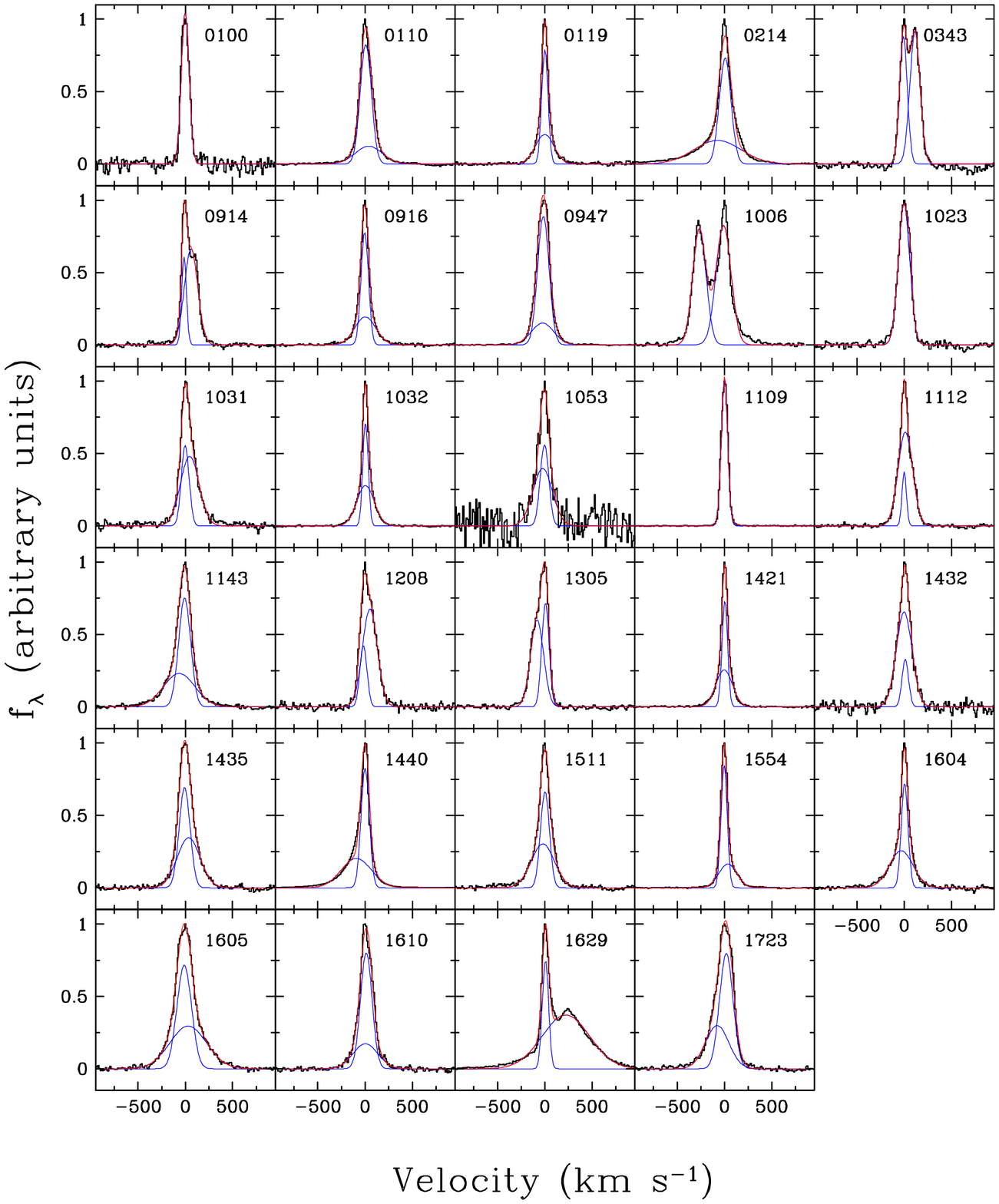}
\caption{[O III] $\lambda5007$ profiles with single or double-Gaussian
  model fits. Individual Gaussian components are shown in blue and the
  summed model is in red.
\label{o3}}
\end{figure}

\begin{figure}
\begin{center}
\scalebox{0.5}{\rotatebox{-90}{\includegraphics{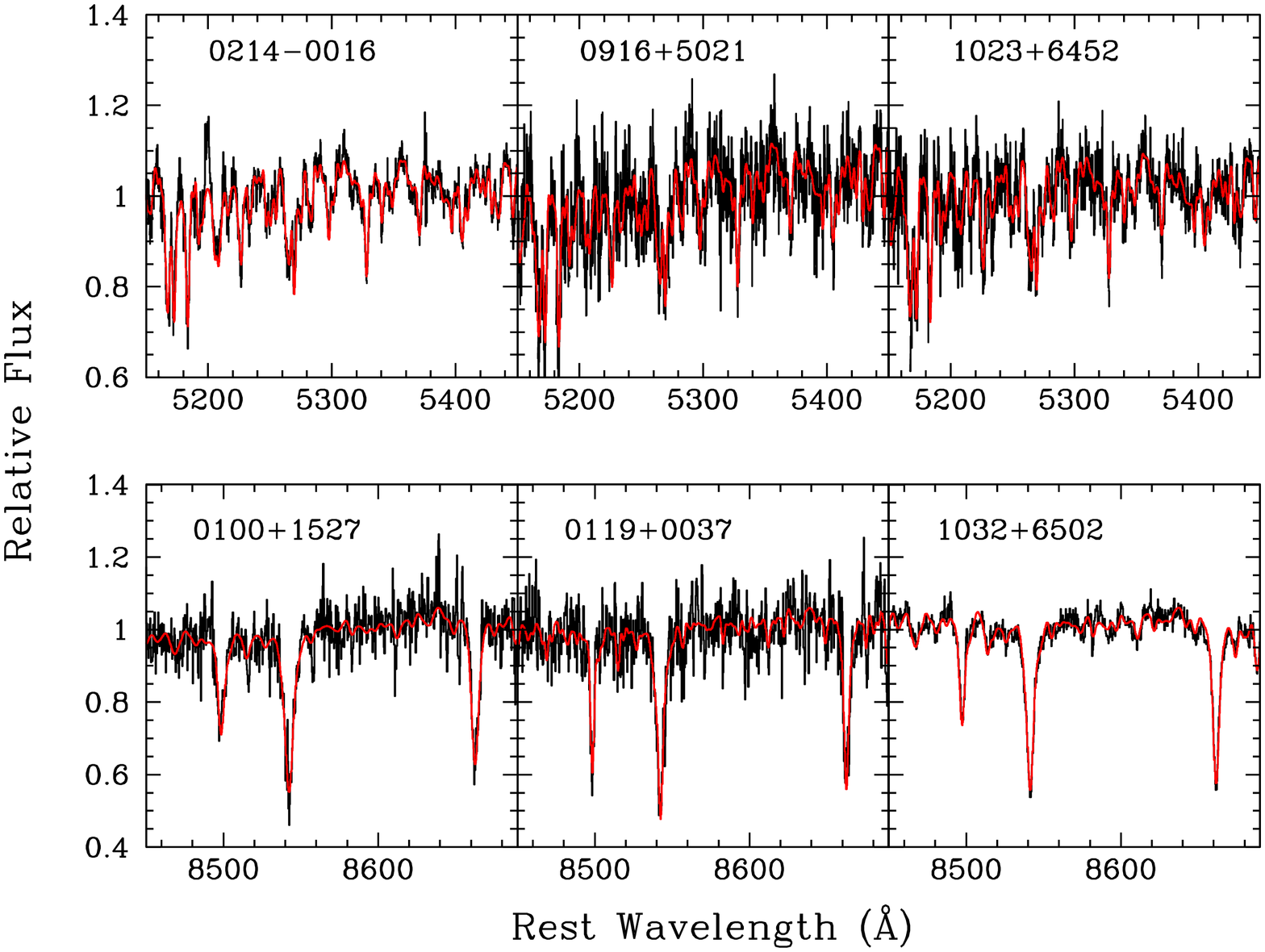}}}
\end{center}
\caption{Examples of fits from velocity dispersion measurements,
  illustrating the range of data quality for the Keck observations.
  Upper panels show fits to the \mgb\ spectral region and lower panels
  show the \ion{Ca}{2} triplet region.  The spectrum in red is the
  best-fitting broadened K-giant template star.
\label{dispersions}
}
\end{figure}

\begin{figure}
\plotone{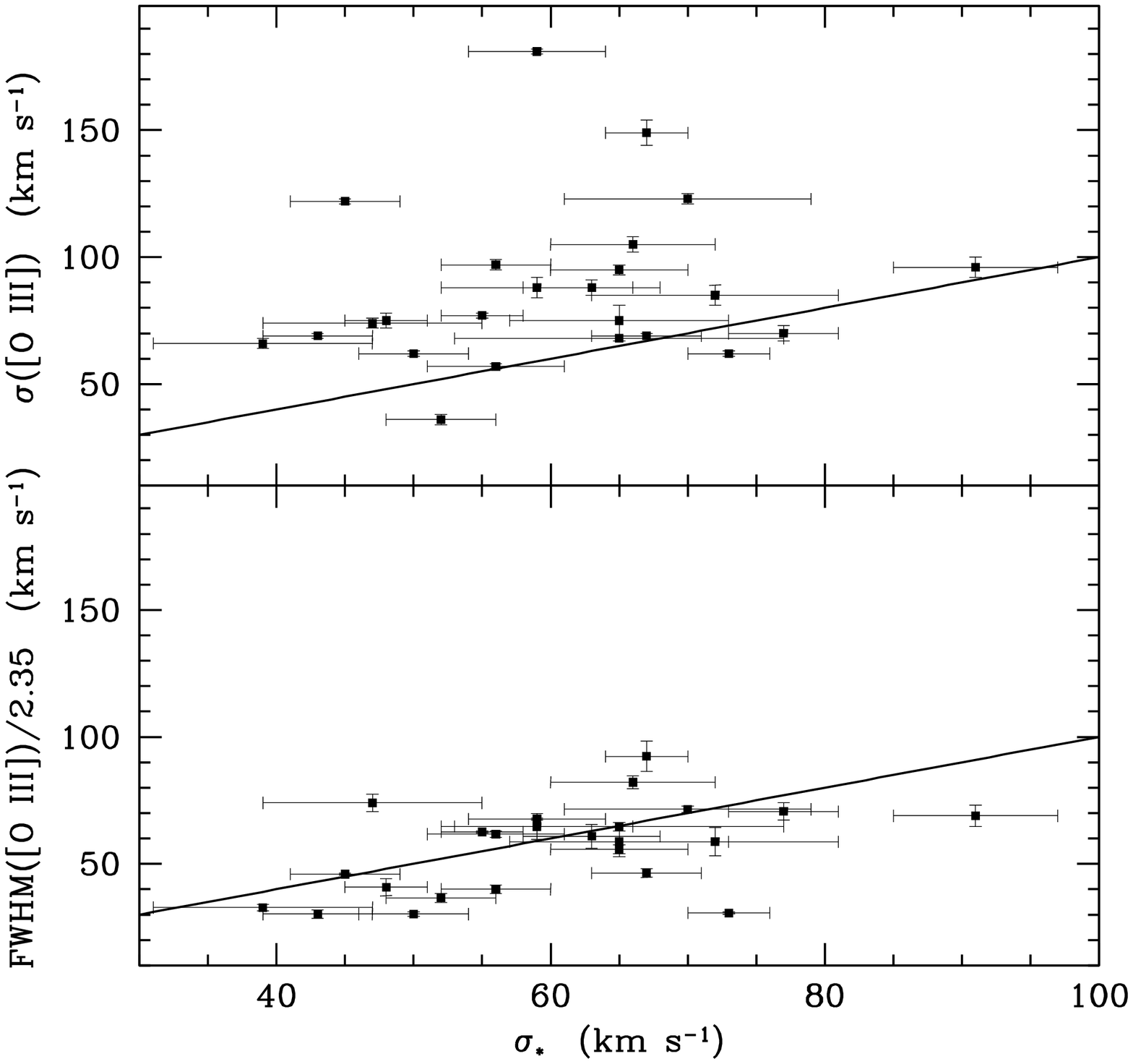}
\caption{Comparison of stellar and emission-line velocity dispersions.
  The upper panel shows the line dispersion \sigmao, and the lower
  panel shows \fwhmo/2.35.  The solid line in each panel represents a
  1:1 correspondence between stellar and gas linewidths.  Objects with
  double-peaked profiles are not included.
\label{linewidths}
}
\end{figure}

\clearpage

\begin{figure}
\begin{center}
\scalebox{0.3}{\includegraphics{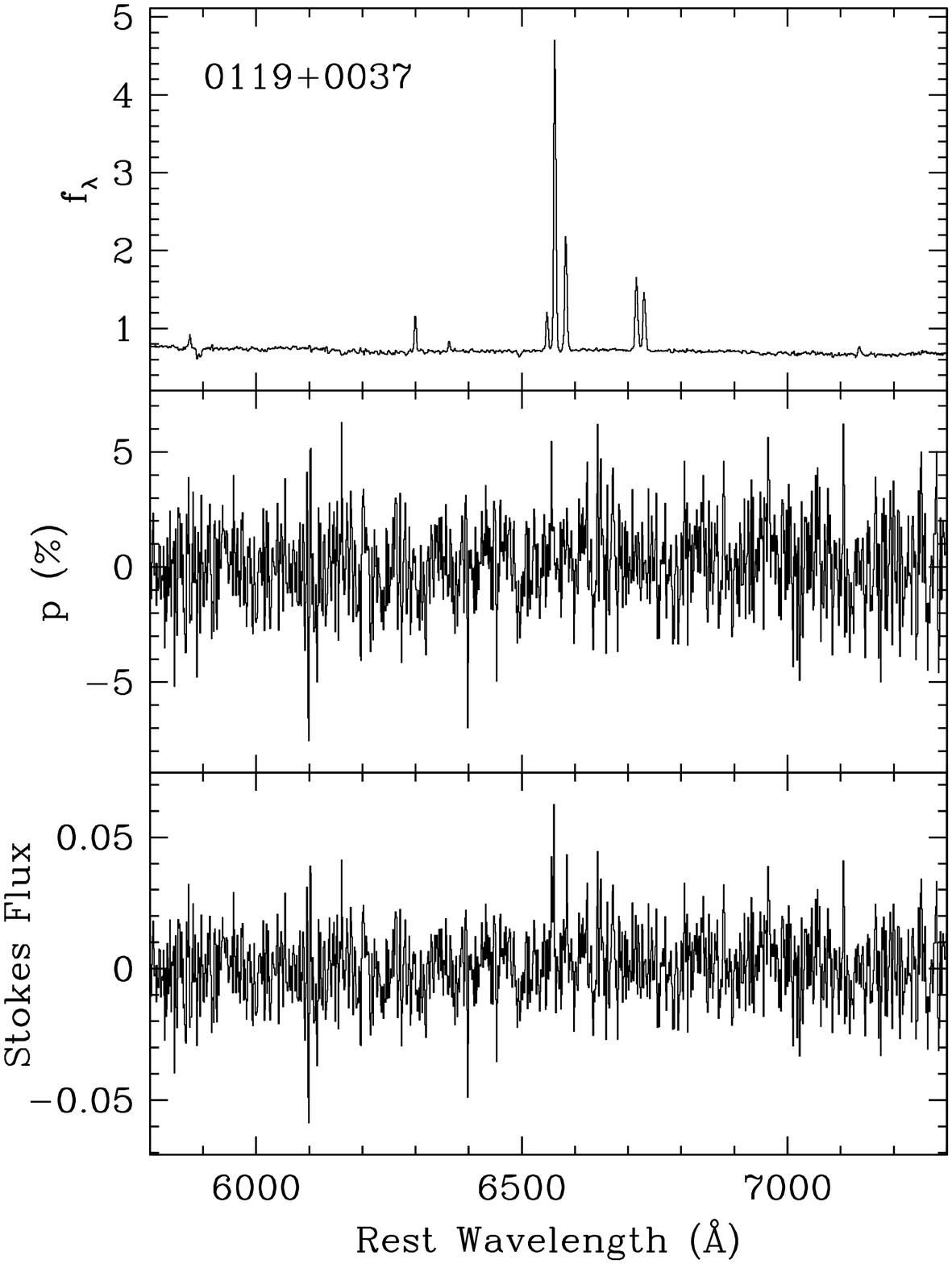}}
\hspace*{0.3in}
\scalebox{0.3}{\includegraphics{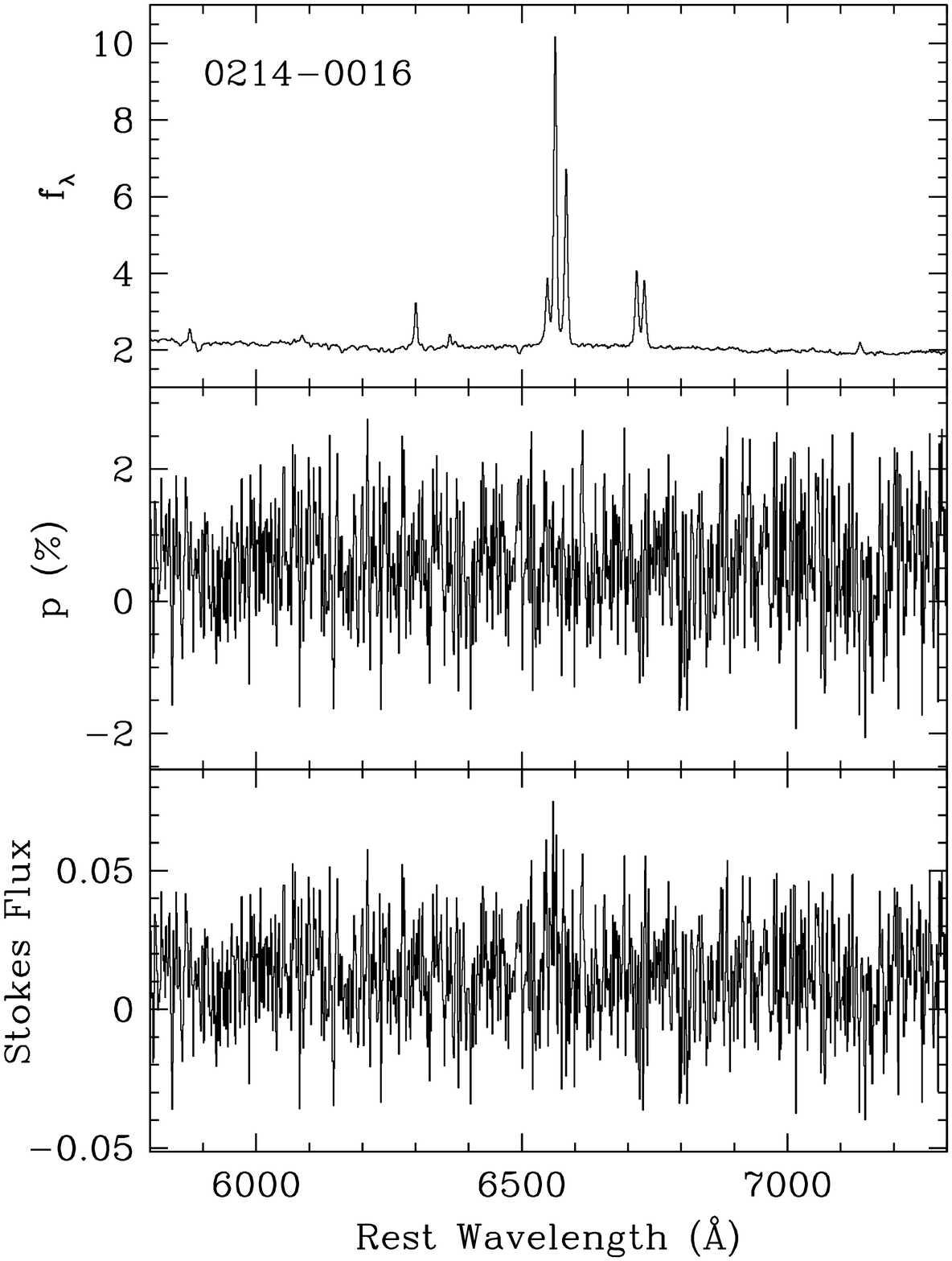}}

\scalebox{0.3}{\includegraphics{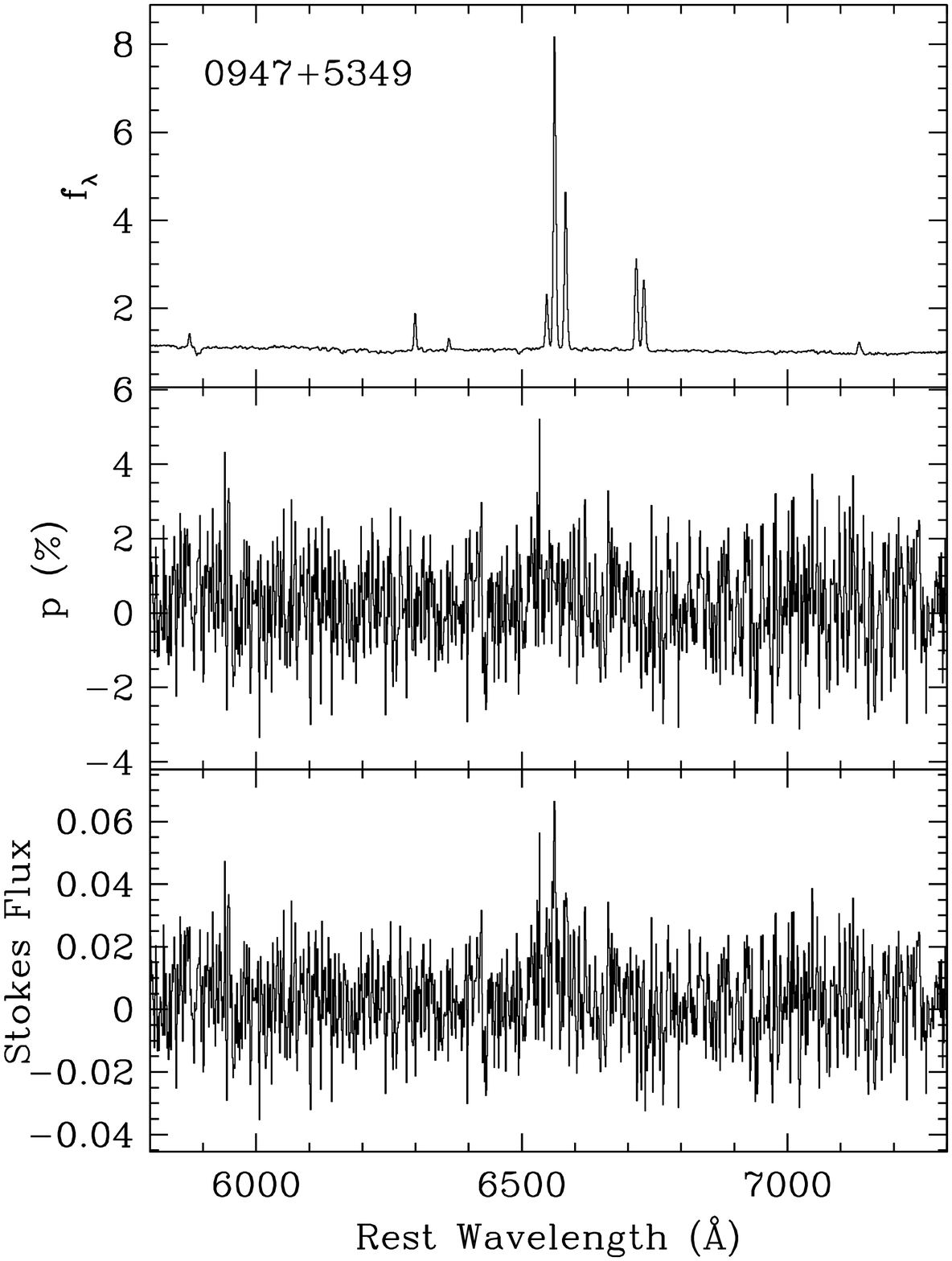}}
\hspace*{0.3in}
\scalebox{0.3}{\includegraphics{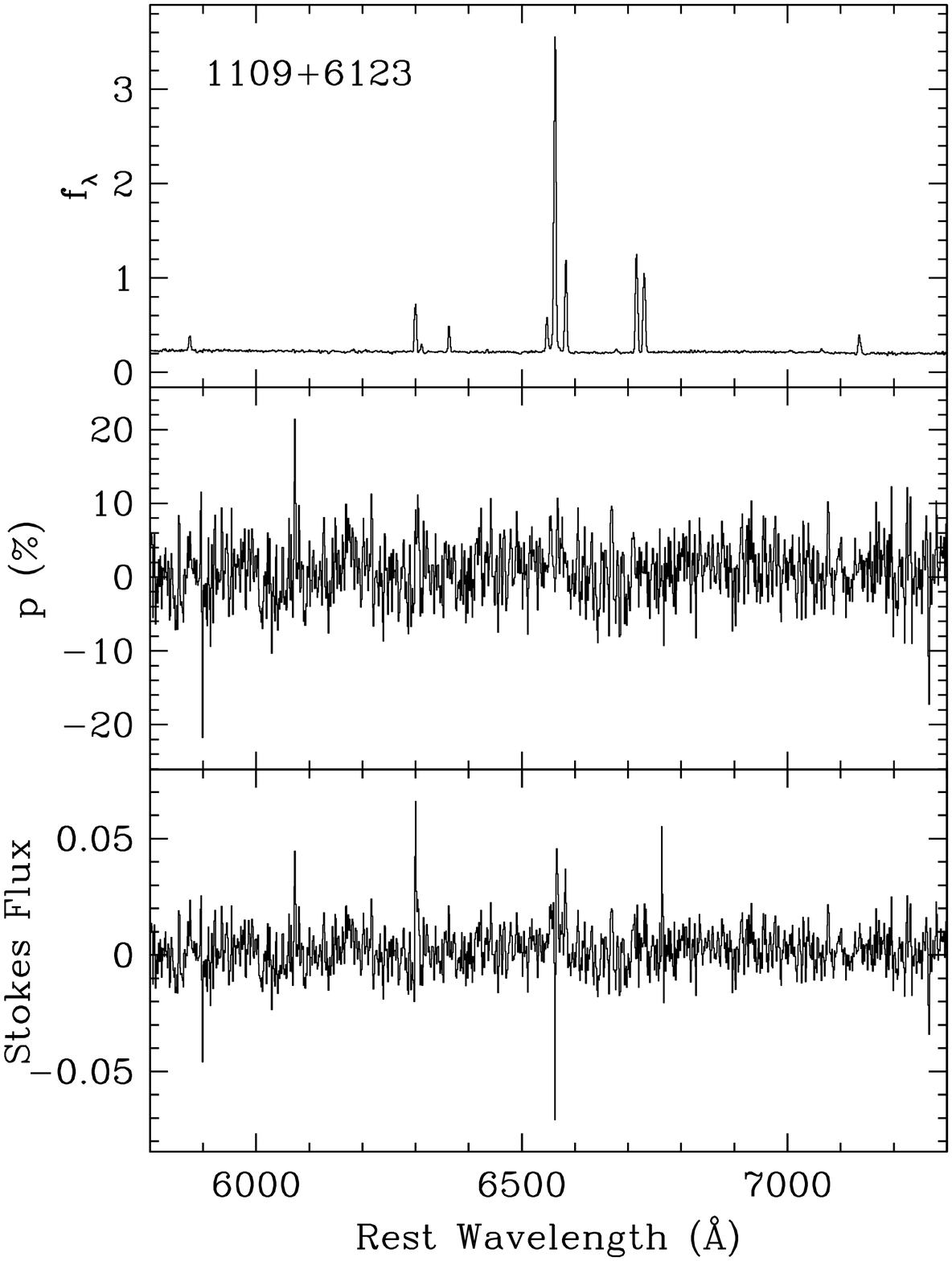}}
\end{center}
\caption{Keck LRIS polarimetry data for four objects in the spectral
  region surrounding the \hal\ emission line.  In each plot, the top
  panel is the total flux spectrum in units of $10^{-16}$ erg s\per\
  cm\persq\ \AA\per, the middle panel is the percent polarization
  given in the form of the rotated Stokes parameter, and the bottom
  panel is the Stokes flux $p \times f_\lambda$.
  \label{polplot}
}
\end{figure}

\begin{figure}
\plotone{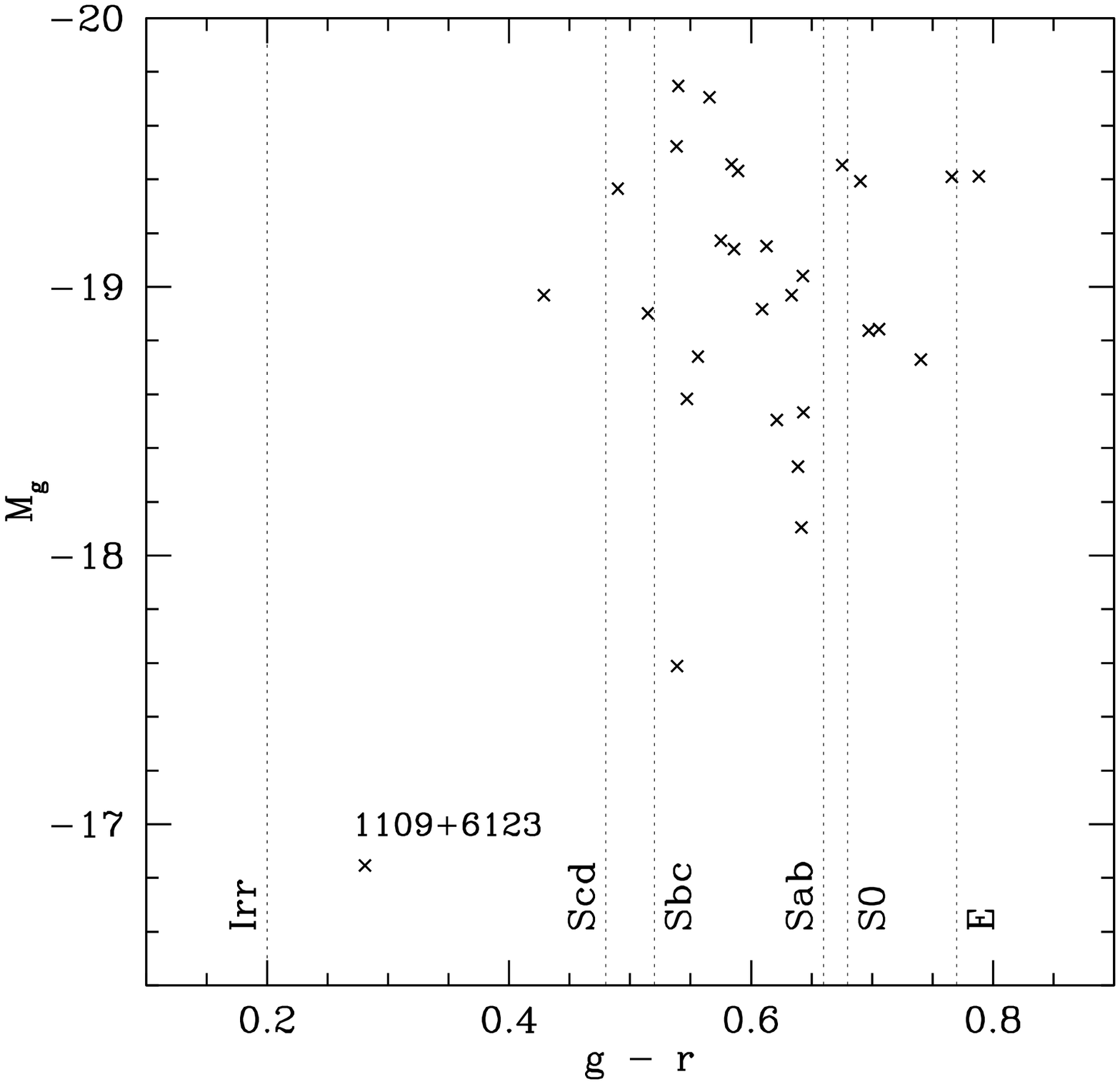}
\caption{Color-magnitude diagram for the galaxies in this sample,
  based on SDSS photometry.  Standard colors for different Hubble
  types are from \citet{fuku95}.
  \label{cmd}
}
\end{figure}

\begin{figure}
\plotone{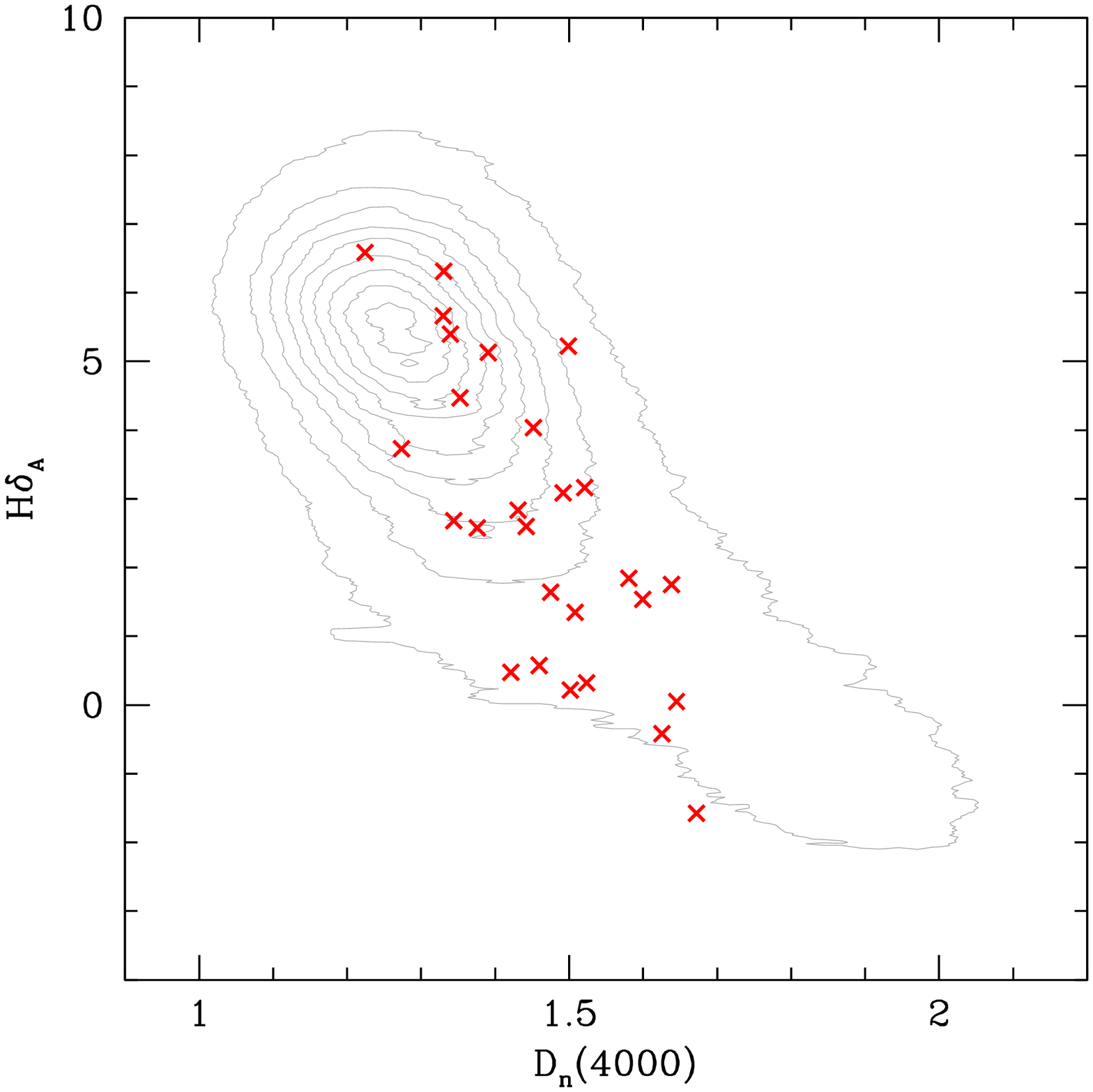}
\caption{The \dn\ and \hdeltaa\ indices for this sample (red crosses).
  Grey contours and small dots represent the distribution of galaxies
  from the \citet{kau03agn} sample within the mass range $9.0 <
  \log(M/\msun) < 10.0$.
\label{hostage}
}
\end{figure}

\begin{figure}
\plotone{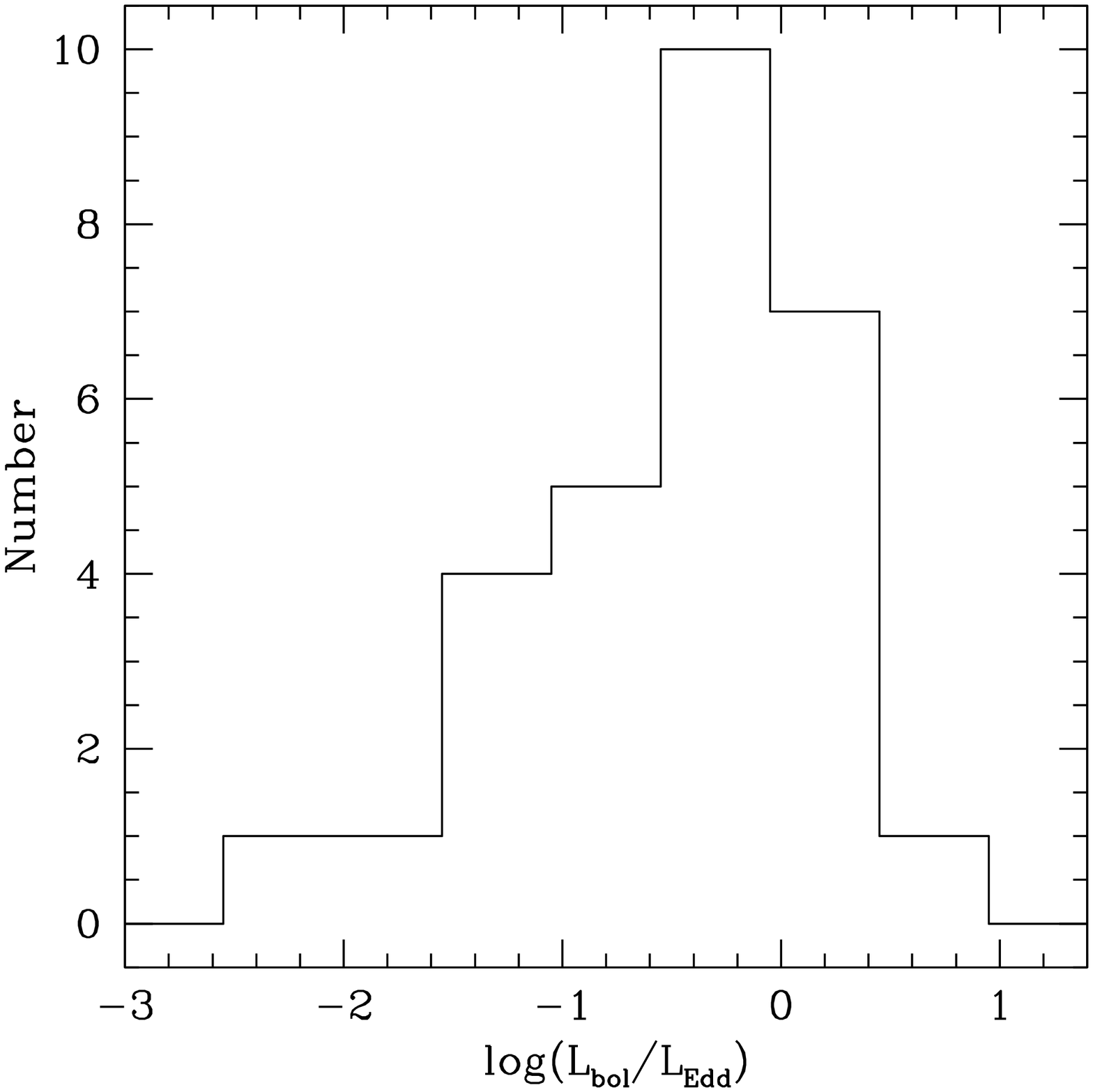}
\caption{Eddington ratio $\lbol/\ledd$ for our sample, based on the
  assumptions of an [\ion{O}{3}] bolometric correction of
  $\lbol/L$([\ion{O}{3}])$=3500$ \citep{hec04} and an offest of 0.23
  dex from the \citet{tre02} \msigma\ relation.
\label{lbol}
}
\end{figure}

\begin{figure}
\plotone{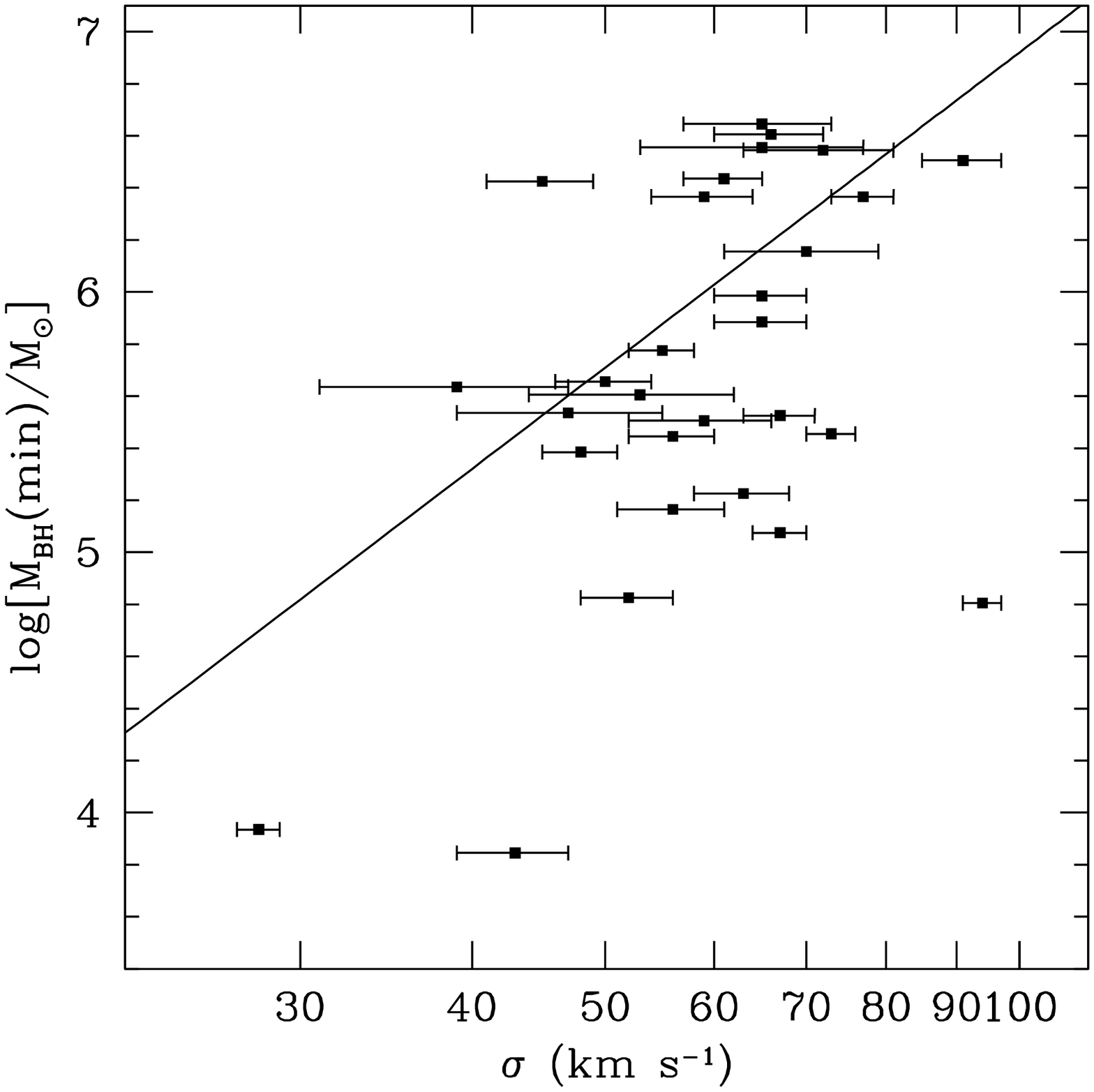}
\caption{Minimum black hole mass vs. stellar velocity dispersion,
  assuming that the black holes are radiating at $\lbol/\ledd<1$, and
  assuming a bolometric correction of $\lbol/L$([\ion{O}{3}])$=3500$.
  For the two galaxies lacking measurements of \sigmastar, the
  quantity FWHM([\ion{O}{3}])$/2.35$ has been used in place of the
  stellar velocity dispersion.  The solid line is the \msigma\
  relation from \citet{tre02}.
\label{bhmin}}
\end{figure}

\begin{figure}
\begin{center}
\rotatebox{-90}{\scalebox{0.5}{\includegraphics{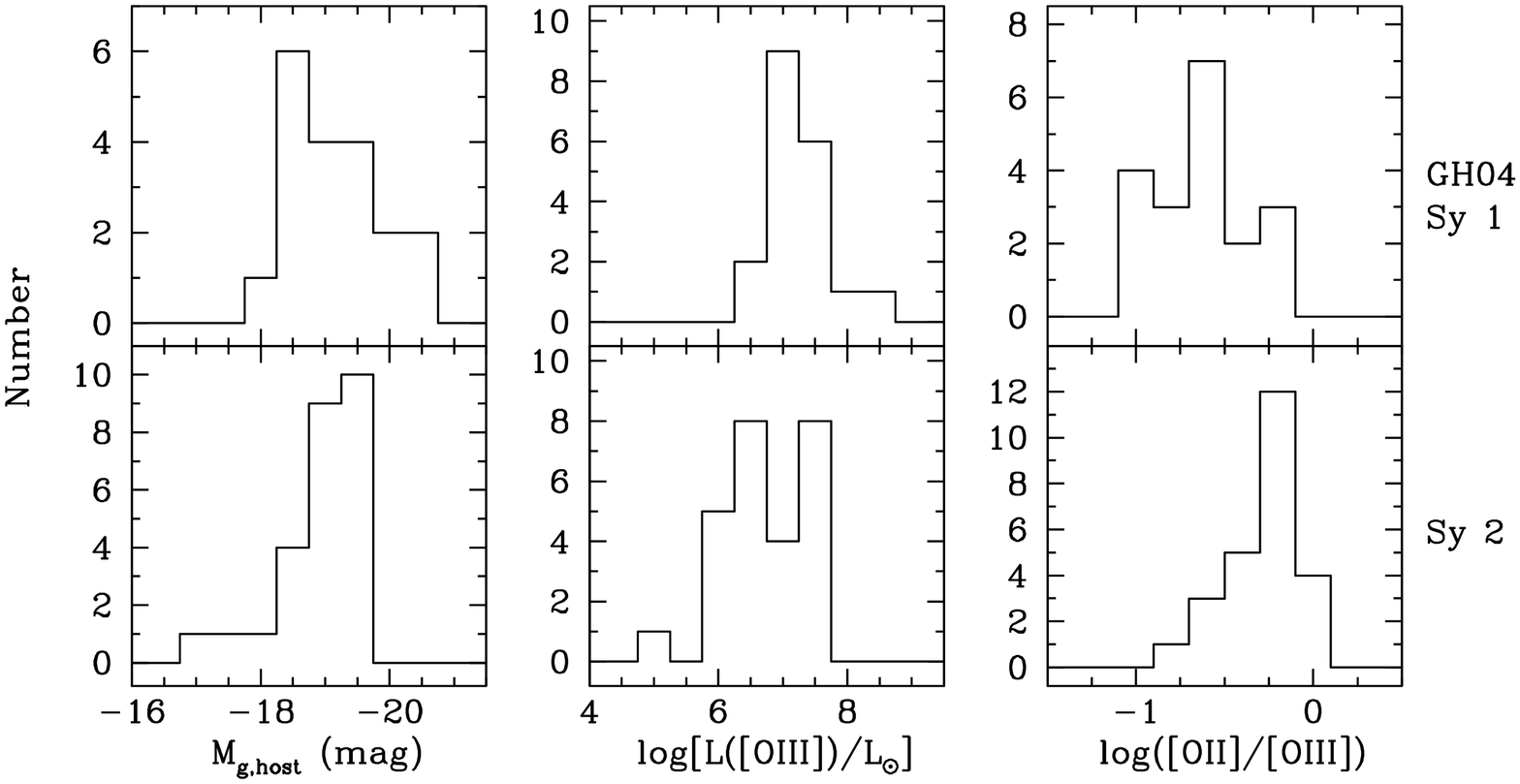}}}
\end{center}
\caption{Histograms of host galaxy absolute magnitude $M_g$,
  [\ion{O}{3}] luminosity, and $\log($[\ion{O}{2}] $\lambda3727 / $
  [\ion{O}{3}] $\lambda5007)$ for the \citet{gh04} Seyfert 1 sample
  (upper panels) and the Seyfert 2 sample described in this paper
  (lower panels).
\label{hostcompare}
}
\end{figure}

\end{document}